\documentstyle[11pt,aaspp4]{article}
\def\lineunits{ergs\ s$^{-1}$\,cm$^{-2}$}
\def\contunits{ergs\ s$^{-1}$\,cm$^{-2}$\,\AA$^{-1}$}
\def\mone{\phantom{1}}

\def\kms{\ifmmode {\rm km\ s}^{-1} \else km s$^{-1}$\fi}
\def\Msun{\ifmmode M_{\odot} \else $M_{\odot}$\fi}
\def\Lsun{\ifmmode L_{\odot} \else $L_{\odot}$\fi}
\def\qo{\ifmmode q_{\rm o} \else $q_{\rm o}$\fi}
\def\Ho{\ifmmode H_{\rm o} \else $H_{\rm o}$\fi}
\def\ho{\ifmmode h_{\rm o} \else $h_{\rm o}$\fi}

\def\gtsim{\raisebox{-.5ex}{$\;\stackrel{>}{\sim}\;$}}
\def\vFWHM{\ifmmode v_{\mbox{\tiny FWHM}} \else
            $v_{\mbox{\tiny FWHM}}$\fi}
\def\CCF{\ifmmode F_{\it CCF} \else $F_{\it CCF}$\fi}
\def\ACF{\ifmmode F_{\it ACF} \else $F_{\it ACF}$\fi}
\def\Halpha{\ifmmode {\rm H}\alpha \else H$\alpha$\fi}
\def\Hbeta{\ifmmode {\rm H}\beta \else H$\beta$\fi}
\def\Hgamma{\ifmmode {\rm H}\gamma \else H$\gamma$\fi}
\def\Hdelta{\ifmmode {\rm H}\delta \else H$\delta$\fi}
\def\Lya{\ifmmode {\rm Ly}\alpha \else Ly$\alpha$\fi}
\def\Lyb{\ifmmode {\rm Ly}\beta \else Ly$\beta$\fi}

\def\heii{He\,{\sc ii}}

\def\ciii{\ifmmode {\rm C}\,{\sc iii} \else C\,{\sc iii}\fi}
\def\civ{\ifmmode {\rm C}\,{\sc iv} \else C\,{\sc iv}\fi}

\def\oiii{O\,{\sc iii}}
\def\o5007{[O\,{\sc iii}]\,$\lambda5007$}

\def\feii{Fe\,{\sc ii}}

\lefthead{Peterson et al.}
\righthead{Variability in Seyfert 1 galaxies}

\begin{document}
\title{Optical Continuum and 
Emission-Line Variability of Seyfert 1 Galaxies}

\author{
Bradley M. Peterson, 
Ignaz Wanders,\altaffilmark{1}
Ray Bertram,\altaffilmark{2}
James F.\ Hunley,
Richard W.\ Pogge,
and R.\ Mark Wagner\altaffilmark{2}
}
\altaffiltext{1}
           {Present address: 
            School of Physics and Astronomy, University of St. Andrews,
            North Haugh, St.\ Andrews, Fife, KY16 9SS, Scotland. Email:
	   iw2@st-and.ac.uk}
\altaffiltext{2}
           {Postal address: 
            Lowell Observatory, 1400 West Mars Hill Road, 
            Flagstaff, AZ 86001. Email: rayb, rmw@lowell.edu} 

\affil{Department of Astronomy, The Ohio State University,\\
    174 West 18th Avenue, Columbus, OH 43210-1106\\
Email: peterson, jhunley, pogge@astronomy.ohio-state.edu}

\begin{abstract}
We present the light curves obtained during an eight-year program
of optical spectroscopic 
monitoring of nine Seyfert 1 galaxies: 3C 120, Akn 120, Mrk 79, 
Mrk 110, Mrk 335, Mrk 509, Mrk 590, Mrk 704, and Mrk 817. 
All objects show significant variability in both the continuum and
emission-line fluxes. We use cross-correlation analysis
to derive the sizes of the broad \Hbeta-emitting regions based
on emission-line time delays, or lags. We successfully measure time delays
for eight of the nine sources, and find values
ranging from about two weeks to a little over two months.
Combining the measured lags and widths of the variable parts of
the emission lines allows us to make virial mass estimates for
the active nucleus in each galaxy. The virial masses are in the
range $10^{7-8}$\,\Msun.
\end{abstract}

\keywords{galaxies: active --- galaxies: Seyfert}
 
\setcounter{footnote}{0}

\section{Introduction}

It has been known since the late 1960s that both
the continuum (e.g., Fitch, Pacholczyk, \& Weymann 1967) and 
broad emission lines (e.g., Andrillat \& Souffrin 1968) 
in active galactic nuclei (AGNs) vary in flux with time.
In the 1980s, spectroscopic monitoring programs showed that
the continuum and emission-line variations are closely coupled,
confirming that the emission-line regions are powered predominantly
by photoionization by the central source. However, the typical emission-line
response times were found to be surprisingly short compared to the
light-travel times expected by most photoionization equilibrium
models (e.g., Peterson et al.\ 1985). Attempts were made to
determine light-travel times for the broad-line region (BLR) by
cross-correlation of continuum and emission-line fluxes
(e.g., Gaskell \& Sparke 1986), but the analyses were plagued by
sparsely sampled light curves and relatively large uncertainties in
the measured fluxes (e.g., Gaskell \& Peterson 1987;
Edelson \& Krolik 1988). In spite of the difficulties, the potential
for extracting physical information about the central regions of
AGNs from variability was generally regarded to be enormous
(see Peterson 1988 for a review of the early monitoring
programs and their implications): in
principle, it is possible to constrain significantly the
structure and kinematics of the BLR by determining the 
emission-line response to continuum variations as a function of
wavelength, since the broad-lines are well-resolved in radial
velocity even at low ($\sim10$\,\AA) resolution. This process is
known as ``reverberation mapping'' (Blandford \& McKee 1982).

Late in the 1980s, it became possible to obtain the quality and quantity of
data necessary to determine emission-line response times.
In the ultraviolet, large amounts of {\it International Ultraviolet
Explorer}\ time were devoted to AGN variability projects
(e.g., Clavel et al. 1991). In the optical, CCDs became widely
available on even moderate-size ($\sim2$-m) and small-size ($\sim1$-m)
ground-based telescope, making it possible to obtain high
signal-to-noise ratio ($S/N$) spectra of high photometric
accuracy with relative ease. The problem of poor time sampling
was obviated by cooperation between observers, either by using many telescopes
(e.g., Peterson et al.\ 1991) or by a group using
a single facility (e.g., Maoz et al.\ 1990, Robinson 1994). Progress in
reverberation mapping through 1992 is reviewed by Peterson (1993),
and a more recent review is given by Netzer \& Peterson (1997).

In 1988, we began a long series of approximately weekly
spectroscopic monitoring of nearby bright Seyfert galaxies
with a CCD spectrograph on the 1.8-m Perkins Telescope at Lowell Observatory.
We present here the first analysis of most of these data.
The scientific goals of the program have been:
\begin{enumerate}
\item To acquire optical continuum and emission-line
light curves of sufficient sampling and quality 
to determine accurately the emission-line
response times, or ``lags'' for a number of AGNs.
\item To investigate AGN continuum behavior over a long
temporal baseline.
\item To investigate the nature of broad emission-line
{\em profile}\ variability, and see what this reveals about
the kinematics of the broad-line region.
\item To investigate the possibility of structural changes
in the BLR on time scales of years (which corresponds to the
dynamical time for the BLR).
\end{enumerate}
Some of the results of this program have been reported
elsewhere in the literature, but this is the first comprehensive
presentation of the data obtained since 1988.

In this paper, we present the light curves for the optical
continuum and the broad \Hbeta\ emission lines in nine
Seyfert galaxies. For most of these objects, we are able to
determine accurately the emission-line lags, and for
these objects we can estimate the mass of the central black hole.
In future papers, we will discuss other issues, such
as line-profile variations.

In \S\,2 we describe the observations and data reduction that led to this
homogeneous data base of spectra. Analysis of the light curves is described
in \S\,3, and we summarize our results in \S,4.

\section{Observations and Data Reduction}

\subsection{Sample Selection}

The galaxies observed in this monitoring program were non-rigorously
selected according to a number of simple criteria. These are
among the brightest Seyfert 1 galaxies observable from the
northern hemisphere. They are distributed in fairly evenly right ascension
since the monitoring program was scheduled approximately
one night per week throughout the entire year, and sources were
observed whenever they were accessible. Many of these galaxies
were found to have variable emission lines based on
earlier observations (e.g., Peterson, Crenshaw, \& Meyers 1985,
and references therein) with the Ohio State Image Dissector Scanner
(Byard et al.\ 1981). Some preference was given to galaxies 
in which the the \Hbeta\ equivalent width had been observed to change, 
which was taken
to be a possible indication that light travel-time effects might
be important in these sources. Some consideration was also given
to ensuring that a wide variety of broad-line profile types
was represented in the sample.

In this paper, we report on results for the nine Seyfert galaxies
listed in Table 1. The common source name
is given in column (1), and the epoch 1950 right ascension and
declination appear in columns (2) and (3), respectively.
The redshift $z$ is given in column (4), and the Galactic
$B$-band extinction $A_B$, from the NED database\footnote{The
NASA/IPAC Extragalactic Database (NED) is operated by the Jet
Propulsion Laboratory, California Institute of Technology,
under contract with the National Aeronautics and Space
Administration.}, is shown in column (5). Column (6)
gives the specific luminosity at about 5100\,\AA\ in the
rest frame of each galaxy, corrected for Galactic extinction,
based on the average fluxes given later in this paper
(Table 5).

In addition to the sources listed in Table~1,
a number of other Seyfert galaxies were monitored spectroscopically
as part of this program, often in coordination with other
observers (particularly the International AGN Watch,
Alloin et al.\ 1994)
and often at multiple wavelengths. The well-studied
source NGC 5548 has been one of our primary targets
(Peterson et al.\ 1991, 1992, 1994, Korista et al.\ 1995),
and recent observations are being prepared for publication.
This program has also included observations of 
NGC 3783 (Stirpe et al.\ 1994),
NGC 4151 (Kaspi et al.\ 1996a),
NGC 7469 (Collier et al.\ 1998),
and the broad-line radio galaxy 3C 390.3 (Dietrich et al.\ 1998),
as well as Mrk 279 and NGC 4051 (to be published elsewhere).
We have published preliminary results on two sources included here,
Mrk 590 (Peterson et al.\ 1993) and Mrk 335 (Kassebaum et al.\ 1997);
the results presented here include revisions of the 
data in these earlier papers, superseding the previous results.
Our observations of another of these sources, Mrk 509,
were included in a larger compilation (Carone et al.\ 1996), and
here we present recent observations obtained since that
program was completed. Our analysis incorporates data
from Carone et al.\ (1996) as well.

Our sample does not include certain well-known nearby very low-luminosity
AGNs; even at the time this program was
initiated (1988), existing data suggested that weekly observations
would seriously undersample the variations in these objects.
Some of these lower-luminosity objects were monitored by
the LAG consortium (see Robinson 1994 for a compilation).
Except for an intensive 10-day program in 1993 (Kaspi et al.\ 1996a),
the best-known Seyfert 1, NGC 4151, has not been included 
as a priority target for this very reason.

We have also deliberately avoided higher luminosity sources
for two reasons:
\begin{enumerate}
\item At the time this program was started,
almost nothing was known about continuum and emission-line variability in
high-luminosity non-blazar AGNs. The scientific focus of this program has
been use of emission-line variability to probe the inner structure of
AGNs rather than study of variability {\it per se}. We therefore
decided to restrict our program to sources that we felt had the best chance of
yielding useful variability data.
\item Higher-luminosity sources tend to be at large redshifts. Inclusion of
high-redshift sources in our program results in observations inefficiencies
as the spectrograph grating angle has to be reset and the instrument
recalibrated for higher-redshift sources. This was deemed to be
undesirable.
We did briefly explore the possibility of including somewhat 
higher-luminosity sources, and in early 1991, we observed GQ Comae
approximately once per week for about three months. No continuum or
emission line variations were detected at that time. Since then,
however, it has been demonstrated that higher-luminosity AGNs
do indeed undergo continuum and emission-line variations 
similar to those seen in the Seyfert galaxies discussed here
(e.g., Kaspi et al.\ 1996b).
\end{enumerate}

The reader is thus cautioned that the sources in Table~1
do not represent a statistical sample on any basis. They were all
selected for this program because we believed that the continuum and
emission-line variability time scales were appropriate for one-week sampling
and because they were readily accessible in terms
of position, redshift, and brightness.
The sources in Table~1 span a mere factor of four
in luminosity, while the AGN phenomenon extends over some 15 magnitudes.

\subsection{Observations and Data Reduction}

All the observations were made with the Ohio State University CCD
spectrograph on the 1.8-m Perkins Telescope of the Ohio Wesleyan and
Ohio State universities at Lowell Observatory near Flagstaff, Arizona.
Observations were scheduled approximately once per week, year round,
interrupted only by bad weather and occasional equipment failures.
Each galaxy was observed whenever it was accessible, although
special priority was assigned to NGC 5548 (discussed elsewhere),
Mrk 335, Mrk 590, Akn 120, Mrk 79, and
Mrk 509. Three of the
sources discussed here, Mrk 110, Mrk 704, and Mrk 817, were
added to the program relatively late, and Mrk 704 and 3C 120
were lower-priority sources, so there are relatively
fewer observations of these galaxies.

The entrance slit of the spectrograph was 
set to a fixed projected width of 5\farcs0,
and a projected  extraction width of 7\farcs6 was used. The large aperture
was used to minimize seeing-dependent aperture effects
(Peterson et al.\ 1995).
The position angle of the slit was kept fixed at 90\arcdeg. A
grating ruled at 350 lines mm$^{-1}$ was used, giving a dispersion of 
$\sim$2\,\AA\ per pixel and a resolution of $\sim$10\,\AA. During
this program, 
two different CCDs were used: through 1991, a Texas Instruments model 4849
chip with 384$\times$592 22\,$\mu$m pixels, and from 1992 onward, a Tektronix
512$\times$512 chip with 27\,$\mu$m pixels. The wavelength range covered
by the observations was somewhat larger than the 4600--5400\,\AA\ range
used in this paper. Most observations were comprised of multiple
(usually three) 20-minute integrations.

The bias subtraction, flat-field correction, wavelength calibration,
and standard-star flux calibration were done in the
standard way using the IRAF data-reduction package. The wavelength 
calibration was based on either Fe--Ne or He--Ar discharge-tube spectra.
Cosmic-ray hits were identified and removed to the fullest possible
extent by comparison of individual exposures.

\subsection{Absolute Flux Calibration}

Absolute spectrophotometry of faint sources is a difficult task, as changes
in atmospheric transparency and seeing highly influences the number of 
photons entering the spectrograph entrance aperture. Only rarely
are conditions identical for observations of program objects and
standard stars, and as a rule, spectrophotometric accuracy better than
10\% is rarely achieved with ground-based observations. Seeing and
transparency variations have weak wavelength dependence, however, so
standard calibration techniques yield accurate {\it relative} 
spectrophotometry. In the case of Seyfert galaxies, {\it absolute}
photometric accuracy can be achieved by using the narrow emission
lines as internal flux standards, since these 
generally do not vary on the time scales of interest 
to us. The narrow emission lines arise in a spatially extended
low-density region; long light-travel 
and recombination times (both typically 100--1000 years)
ensure that variations on much shorter time scales will not occur.

The absolute flux scale in each case is established by
measuring the [\oiii]\,$\lambda5007$ flux in the minority of
spectra that were obtained under photometric conditions, as
judged by the observer on site and at the data reduction
stage, where individual exposures were compared. In 
Table~2, we give the absolute fluxes 
that we have determined for the nine Seyfert galaxies
in this study.

A potential problem with this method of flux calibration is
that seeing-dependent aperture effects can be important,
particular if the narrow-line region is spatially resolved
and comparable in size to the projected entrance aperture
of the instrument. It is in principle possible to correct
for such effects by simulated aperture photometry of
emission-line and host-galaxy images (Wanders et al.\ 1992; 
Peterson et al.\ 1995). However, none of the objects here
has extended emission of sufficient strength to affect any
of our results.

Once the absolute [\oiii] flux is determined for each source,
we scale each spectrum by a multiplicative constant so
that the [\oiii] flux in each spectrum of a given object is
the same. We carry this out by comparing each spectrum to
a high signal-to-noise ``reference'' spectrum that is
constructed by averaging all of the highest-quality spectra
and scaling this to the correct absolute flux. 
The scaling is accomplished by using 
the automatic scaling program developed by van Groningen \& Wanders (1992),
which performs the task in a fast, reliable, and objective way, and has
been successfully used on a number of data sets.
The program also corrects for small zero-point
wavelength-calibration errors between the
individual spectra, and takes resolution differences into account. The
resulting set of calibrated spectra
is therefore highly homogeneous.  A few of the spectra were poorly focused,
which made flux calibration very difficult and the results
extremely suspect. These spectra were removed from the data base.

All of the acceptable spectra were used to compute 
an average spectrum for each object. For each source, 
we also computed a root-mean-square
(rms) spectrum,  defined as
\begin{equation}
\sigma(\lambda)=\left\{{1\over (N-1)}\sum_{i=1}^N \left[F_i(\lambda)-\bar 
F(\lambda)\right]^2\right\}^{1/2},
\end{equation}
where the sum is taken over the $N$ spectra, and $\bar F(\lambda)$ is the
average spectrum. The rms spectrum is thus a measure of the variations around
the mean, and constant features, such as narrow emission lines, Galactic
absorption lines, and the non-variable parts of the broad emission lines and
the continuum, are eliminated. Any narrow-line residuals visible are due to
imperfections in the data calibration. These are usually very small 
($<1$\% in any given spectrum),
but add up in the rms spectrum, often rendering them visible.
The mean and rms spectra for these objects are shown in
Figs.\ 1--3.

\subsection{Light Curve Measurements}

Once all of the spectra have been calibrated, continuum and emission-line
flux measurements were made. For each object, the continuum flux is
measured in band about 15--20\,\AA\ wide
at around 5100\,\AA\ in the rest frame of the object, as this is 
the most line-free region in this part of the spectrum. 
The exact wavelength limits
used vary from object to object, depending on the strength and width
of various emission lines, especially the \feii\ blends.

The \Hbeta\ emission-line flux is measured in a simple fashion. First,
and underlying continuum is interpolated between the local minima
between \heii\,$\lambda4686$ and \Hbeta\  on short-wavelength side of 
\Hbeta\ and
between \Hbeta\ and the \feii\ blend on the long-wavelength side of
\Hbeta. We then integrate the total flux above this continuum between
the short-wavelength limit and a point just shortward of the
[\oiii]\,$\lambda4959$ line. This measurement includes some
\feii\ emission as well as the narrow-line \Hbeta\ component, and
misses \Hbeta\ flux at high positive radial velocities, i.e., the
extreme redward wing of the feature. However, this simple measurement is
sufficiently good for our current goals, and is
model independent. The wavelength boundaries for this various
integrations are given in Table~3. In each case,
the wavelengths limits given are in the observed frame.
Column~(1) identifies the object, and column (2) gives the
lower and upper bounds of the continuum region. 
The \Hbeta\ integration limits are given in column (3);
the \Hbeta\ flux is taken to be the total flux above
the linear pseudo-continuum defined by the limits given
in column (4). In some, but not all, cases, \heii\ emission can also
be determined, and these measurements will be reported elsewhere. It
is also apparent from some of the rms spectra that some iron
features, such as \feii\,$\lambda5018$ in Mrk 335, Mrk 590, and Mrk 817,
also have varied.

Assigning uncertainties to the light curve measurements is 
not straightforward, and we treat this problem in a heuristic
manner. Most of the spectra have high signal-to-noise ratios
($S/N \gtsim 50$), and for these we assign uncertainties based
on differences between observations that are closely spaced in
time (within a few days of each other). 
This provides a conservative error estimate, since
any real low-amplitude amplitude variability is attributed to
random error; the uncertainties 
may thus be somewhat smaller than those we assign by this method.
We assign fractional errors of 0.02 to continuum and emission-line
measurements.

Some of the spectra, however, are of lower quality, usually on
account of abnormally short exposure times due to high airmass
(at the beginning or end of an observing season for a particular
source) or on account of poor or highly variable atmospheric
transparency. To assign errors to these spectra, we first
determine an apparent
``signal-to-noise'' ratio for each spectrum by computing the
standard deviation of the flux values in the continuum band
at about 5100\,\AA\ in the rest frame (this is usually 
based on 7--10 pixels). For high quality data, this
apparent $S/N$ is lower than the actual signal-to-noise
ratio of the spectra because weak emission and absorption
structures are counted as random noise.
For spectra with apparent $S/N < 50$,
we assign a fractional error of $1/(S/N)$ to both the continuum
and line measurements. We tested this procedure by 
first producing very high $S/N$ spectra by averaging a large
number of the highest quality spectra of each source, and then
performing Monte Carlo simulations where we 
added random Gaussian-distributed
noise to the spectrum and measured the continuum
and emission-line fluxes in the same fashion as 
the real spectra. A large number of realizations for a given
$S/N$ revealed that the standard deviation of the measured
fluxes was approximately equal to the mean flux divided by
$S/N$. Thus, all continuum and emission-line flux measurements
are taken to be accurate to $\sim2$\%, except those based on
lower-quality data, in which case the fractional errors in
both the continuum and line fluxes are taken to be $1/(S/N)$.

The final light curves for the various objects are shown in 
Figs.\ 4--12. The data can be obtained in tabular form 
through the World-Wide Web\footnote{The light curves 
and complete logs of observation are
available in tabular form, in either PostScript or plain
ASCII format at URL 
{\sf http://www.astronomy.ohio-state.edu/$\sim$peterson/AGN/}.}.
All measurements are in the observer's
frame, and are uncorrected for Galactic extinction.

\section{Analysis and Discussion}

\subsection{Characteristics of the Data Base}

The general sampling characteristics of the light curves
plotted in Figs.\ 4--12 are given
in Table 4. Column (1) gives the source name and column (2)
gives the total number of observations that comprise the light curve.
Column (3) gives the number of days spanned by the observation,
and the average and median intervals between observations are
given in columns (4) and (5), respectively.

The primary targets (Akn 120, Mrk 79, Mrk 335, Mrk 509, and Mrk 590)
are all well-sampled, with median intervals between observations
of 7--8 days for a span of 7--8 years. The larger average intervals between
observations are an effect of seasonal gaps that occur when the sources
are near conjunction and thus unobservable. 
Of the other sources, Mrk 110 and Mrk 817 are
also well-sampled, but for only 4--5 years. Both 3C 120 and Mrk 704
are quite poorly sampled, as will become apparent when we carry out
the time-series analysis.

All of the sources observed underwent significant continuum and
emission-line variations. Statistics that describe the 
variations are given in Table 5. For each object, we give
for both the continuum and \Hbeta, 
the average flux $\langle F \rangle$ (columns 2 and 5),
the normalized variability amplitude $F_{var}$, i.e., the rms
fractional variability corrected for measurement uncertainties,
as defined by Rodr\'{\i}guez-Pascual
et al.\ (1997) (columns 3 and 6), and 
the ratio of maximum to minimum flux $R_{max}$ (columns 4 and 7).
The average continuum
flux reported here is the value used to compute the 
luminosity given in Table 1. The variability parameters have
not been corrected for the constant contaminants, starlight
in the case of the continuum, and the \Hbeta\ narrow-line
components in the case of the line. 
These issues are beyond the scope of this contribution
and will be dealt with in the future.

\subsection{Time-Series Analysis}

The primary goal of this program has been to determine for each of
these sources the time delay, or lag, between the continuum
and \Hbeta\ flux variations. The existence of time delays is in most
cases apparent from close inspection of the light curves.
We quantify these time delays by cross-correlating 
the continuum and emission-line light curves for each object.
Most of the light curves are well-sampled, so the cross-correlation
method that we use is the interpolation cross-correlation function
(ICCF) introduced by Gaskell \& Sparke (1986) and Gaskell \& Peterson
(1987). In cases where there are significant gaps in the data
that may compromise the validity of linear interpolation between
adjacent measurements, we have also employed the discrete-correlation
function (DCF) method (Edelson \& Krolik 1988) as a check on the results.
The specific implementations of the ICCF and DCF that we employ are
described by White \& Peterson (1994).

The uncertainties in the
cross-correlation results are estimated by using the model-independent
FR/RSS Monte Carlo method described by Peterson et al.\ (1998). 
Each Monte Carlo simulation consists two parts, which are
referred to as ``random subset selection'' (RSS) and
``flux randomization'' (FR). The RSS procedure consists of randomly
drawing from a light curve $N$ points a new sample of $N$ points, chosen
without regard to whether any particular point has been previously
drawn. In this regard, RSS is reminiscent of a standard statistical
``bootstrap'', although it differs in that the temporal order of the
points must be  preserved. It is only {\em after}\ $N$ points have been
selected that the redundant selections are removed from the
sample. This effectively reduces the number of points in each light
curve by a factor of $\sim1/e$. The RSS procedure thus 
accounts for the effects that individual data points may
have on the cross-correlation by removing them at random. 
The second part of the procedure (FR) is intended to account
for the effects of flux-measurement uncertainties.
The observed fluxes are altered by
random Gaussian deviates scaled to the uncertainty ascribed to each point.
Peterson et al.\ (1998) demonstrate that under a wide variety of
fairly realistic conditions the combined FR/RSS procedure
yields conservative errors, in the sense that the
real uncertainties may in fact be somewhat smaller than the errors quoted.

Cross-correlation calculations have been carried out using the
entire data set for each source, and in most cases, also using
data from single seasons (usually spanning about 200 days) in
which the variability characteristics were favorable for 
accurate detection of a lag. The only data set that failed to
yield a statistically significant lag estimate was that for
Mrk 704, which is not surprising as this is the most poorly
observed galaxy in our sample.
Cross-correlation functions (CCFs) computed
from each data set in its entirety are shown in Fig.\ 13, which
shows in each case the ICCF (continuous line) and DCF (individual
error bars). In each case, the ICCF and DCF are in excellent
agreement near the peak, which is the area of interest. The ICCF
and DCF often show disagreement for lags of 100 days or more, since
it is in this regime where the large seasonal gaps become important
and the underlying assumptions of the ICCF begin to break down.
The peaks and centroids of the CCFs are, however, generally
in good agreement.

The results of our cross-correlation analysis are presented in Table 6.
Column (1) gives the object name, and the subset of the light curve
used in the calculation is given in column (2). The
ICCF centroid $\tau_{cent}$ is given in column (3). The peak of
the cross-correlation function occurs at a lag
$\tau_{peak}$ (column 4) and has value $r_{max}$ (column 5). The
full-width at half maximum (FWHM) of the ICCF is given in column (6),
and the total number of observations in the light curve appears in
column (7). The centroid $\tau_{cent}$ is computed using all points
with correlation coefficients $r \geq 0.8r_{max}$. The uncertainties
given for $\tau_{cent}$ and $\tau_{peak}$ were computed with the
FR/RSS methodology; a large number ($\sim1000$) of Monte Carlo
realizations were used to build up a cross-correlation peak
distribution (CCPD; Maoz \& Netzer 1989), which is then integrated
to given the uncertainties. The range of uncertainties contain
68\% of the realizations, and thus would correspond to $1\sigma$ uncertainties
for a normal distribution.

Inspection of Table 6 shows that both $\tau_{peak}$ and $\tau_{cent}$
can vary considerably among the various subsets. While this
might plausibly be ascribed to real physical changes in the structure
of the BLR (e.g., Wanders 1995), the differences probably
are simply indicative of a thick BLR (i.e., the ratio of outer
to inner radius is much greater than unity). For a simple
linear model of the line response $L(t)$, we can write
\begin{equation}
L(t) = \int\! \Psi(\tau) C(t-\tau)\,d\tau,
\end{equation}
which is known as the transfer equation, and $\Psi(\tau)$ is the
transfer function (Blandford \& McKee 1982), which depends on
the BLR geometry, viewing angle (inclination to the line of sight),
and the line-reprocessing
physics. By convolving this with the continuum light curve
$C(t)$, it is easy to show that the cross-correlation function
(\CCF) can be written
\begin{equation}
\CCF(\tau) = \int\! \Psi (\tau')\,\ACF(\tau-\tau')\,d\tau',
\end{equation}
and \ACF\ is the continuum autocorrelation function
(Penston 1991; Peterson 1993). The
CCF is thus sensitive to the particulars of the continuum variations.
We performed some simple experiments by convolving the
observed continuum light curves with model transfer functions
for thick spherical shells. We found that it is relatively simple
to obtain a wide range of lags if the geometry is thick and the
response not strongly biased to particular radii. This point was
first made by Netzer \& Maoz (1990) to explain the large differences
in the \Hbeta\ response determined by Peterson et al.\ (1991) and
Netzer et al.\ (1990). Again, this
is a topic that will be pursued more completely elsewhere.

In Table 7, we summarize the cross-correlation results. For
$\tau_{cent}$, we adopt the value that gives the smallest 
uncertainty for each source, and we adopt $c\tau_{cent}$ as
an estimate of the BLR size. In the cases where subsets of
the light curves yield the smallest FR/RSS error estimates,
we show the cross-correlation functions based on the
minimum-error subset in Fig.\ 14. We have also measured the
width of the broad \Hbeta\ line in the rms spectrum
$\vFWHM$, since this provides a measure of the radial-velocity
distribution of the gas that is actually varying. We adopt
as a Keplerian velocity
\begin{equation}
v = \frac{\sqrt{3}}{2} \vFWHM
\end{equation}
(Netzer 1990) and we use this to obtain a virial estimate of the mass of
the central source
\begin{equation}
\label{eq:mass}
M = \frac{v^2 c\tau_{cent}}{G}.
\end{equation}
We give the virial mass for each source in column (4) of
Table 7. These masses are highly uncertain since neither the geometry
nor the kinematics of the BLR are known; indeed, we have assumed
that the BLR velocity field is not primarily radial, which seems
to be generally consistent with the reverberation results for
other galaxies. In any case, eq.\ (\ref{eq:mass}) should contain
a numerical factor of order unity that is depends on the
detailed geometry and kinematics of the BLR. Thus the masses
given in Table 7 are valid only as order-of-magnitude
estimates.

\subsection{Comments on Individual Objects}

We conclude this section with some brief comments on a few of
the objects.

\subsubsection{3C 120}
Continuum and emission-line variability was first reported in
this source by Oke, Readhead, \& 
Sargent (1980) and French \& Miller (1980), who placed an upper
limit on the size of the \Hbeta-emitting region of $\sim0.2$\,pc
$\approx 240$\,light days. 
Our light curve is not particularly well-sampled; however, we
do find a statistically significant lag, but with a large
formal uncertainty ($\tau_{cent} = 43.8^{+27.7}_{-20.3}$\,days).

\subsubsection{Akn 120}
The galaxy Akn 120 is of special interest to us, as it provided
the original evidence that the BLR is more compact than thought
at the beginning of the last decade (Peterson et al.\ 1985).
Through the mid-1980s, we monitored Akn 120 regularly with
the OSU IDS (see Peterson, Korista, \& Wagner 1989, and references
therein). Several problems prevented obtaining an accurate
measurement of the lag: 
\begin{enumerate}
\item The temporal sampling, originally
selected on the assumption that the BLR was a light year or so in
radius, was too poor to resolve the line response.
\item The typical uncertainties in the IDS measurements were 
$\sim$8\%; short time-scale, low-amplitude variations simply could
not be detected.
\item The narrow [\oiii]\,$\lambda\lambda4959$, 5007 lines are
unusually weak in Akn 120 (see Figs.\ 1--3), and flux
calibration based on  [\oiii]\,$\lambda5007$ is less reliable than
it is normally. This leads to systematic ``correlated errors'' in the
continuum and emission-line fluxes at zero lag; i.e., any error
in the [\oiii]\,$\lambda5007$ calibration introduces a systematic
error that drives the continuum and line fluxes in the same
direction, thus artificially enhancing the cross-correlation at $t=0$.
\end{enumerate}
The large flux uncertainties presented the biggest problem with the 
earlier IDS data, since simulations suggested that the 
systematic correlated errors would not matter if the total error
level could be decreased by a factor of two or more (Gaskell \& Peterson 1987).
In fact, replacement
of the IDS spectrograph with the CCD spectrograph has decreased the
uncertainties by a factor of four.

The combined IDS and CCD 
continuum and \Hbeta\ light curves of Akn 120 are shown in Fig.\ 15.
The dramatic improvement in the signal-to-noise of the light curves
between the two instruments is readily apparent. It might seem
disturbing that the amplitude of variability seems to have decreased
with the advent of the improved instrument, but this type of behavior is seen
in more homogeneous data as well. Mrk 509 (Fig.\ 9) shows a similar
episode of violent variability followed by lower-amplitude variations.
Since it was these dramatic variations in Akn 120 that led us directly
to programs such as reported here, we do not regard the difference
between the variations seen in the early 1980s and those in the
early 1990s as especially suspicious.

Figure 16 shows the result of cross-correlating the IDS continuum and
\Hbeta\ light curves (also shown in Fig.\ 17 of Peterson, Korista, \&
Wagner 1989). The centroid of the ICCF shown in Fig.\ 16 is
$\tau_{cent} = 8.7^{+9.9}_{-10.2}$\,days, which is certainly erroneous 
because the systematic errors have not been included in the FR/RSS simulations.
The sharp peak at zero lag demonstrates the strong influence of
correlated errors in this source. Peterson, Korista, \& Wagner point
out that the asymmetric part of this function peaks at $51 \pm 12$\,days,
which is generally consistent with the results reported in Table 6,
based on the higher-quality CCD data. 
The asymmetric peak around zero lag is also found using the DCF method,
also shown in Fig.\ 16. This is 
contrary to what is stated by Edelson \& Krolik (1988), and due
to the fact that the version of the DCF code used by Edelson \& Krolik
contained an error in how the data were weighted
(see White \& Peterson 1994, footnote 2).
Our version of the DCF does not
weight the data at all. Figure 16 should be
compared directly with Fig.\ $5b$ of Edelson \& Krolik.

The cross-correlation lags that we obtain for Akn 120 are also generally
consistent with the estimates of Gaskell \& Peterson (1987)
($\tau = 14 \pm 21$\,days, based on a Monte Carlo model that included the
effects of correlated errors) and  of Peterson \& Gaskell (1991)
($\tau = 39 \pm 14$\,days, based on combined UV/optical data obtained
during the largest outburst).

\subsubsection{Mrk 79}
Markarian 79 has been included on our list of priority sources
since the early 1980s, as our first observations with the OSU IDS
showed it to be one of the highest-amplitude variables in our survey
(Peterson, Crenshaw, \& Meyers 1985). Like Akn 120 and NGC 5548, it
was included in our monthly monitoring program in 1983--85; these
data failed to yield a cross-correlation lag, apparently because the
light curves were badly undersampled (Peterson \& Gaskell 1986).
Oddly enough, over eight years of our CCD-based program, Mrk 79
has been one of the {\em least}\ variable sources in our sample.
The variations were sufficient, however, for a lag to be measured
from the entire data set, as well as in four individual subsets.

\subsubsection{Mrk 110} 
Our interest in Mrk 110 dates back to 1983--84 when we observed an
enormous change in the strength of \heii\,$\lambda4686$ in two spectra
obtained a year apart (Peterson 1988). This behavior is also
apparent in the rms spectrum of Mrk 110 shown in Fig.\ 2.

\subsubsection{Mrk 335}
This galaxy was included in our sample on the basis of an earlier
study by Shuder (1981) that suggested the variations would be well-resolved
by weekly monitoring. A preliminary analysis based on most of the
data presented here was presented by Kassebaum et al.\ (1996). 
The light curves presented here supersede those presented by
Kassebaum et al., but the cross-correlation results are not
changed substantively by using the revised light curves.

\subsubsection{Mrk 509}
Most of the data presented here previously appeared in a large
compilation by Carone et al.\ (1996). The new data
presented here have simply been appended to the light
curves presented by Carone et al. The mean and rms spectra
shown in Fig.\ 2, however, are based {\em only} on the OSU
CCD spectra.

\subsubsection{Mrk 590} 
Markarian 590 was included in our program because our early
IDS observations showed that it underwent enormous \Hbeta\
variations. Additional observations indicated that the emission-line
variations were concentrated in the core of the line; the wings
apparently vary much less, leading Ferland, Korista, \& Peterson (1990)
to suggest that the high radial-velocity \Hbeta\ emission arises
in an optically thin gas. A preliminary version of a subset of our CCD 
observations (JD2448090--2448323) has been previously published
(Peterson et al.\ 1993).

\subsubsection{Mrk 704 and Mrk 817}
We included both of these sources as secondary targets rather late
in the program, based on their position in the sky and because 
they have interesting, complex \Hbeta\ profiles. Mrk 704 was too
poorly sampled to yield a cross-correlation result; nevertheless,
interesting profile variations were observed, as seen in the rms
profiles in Fig.\ 3. These will be discussed elsewhere.

\section{Conclusions}
In this contribution, we have reported on the initial results of
an eight-year spectroscopic monitoring program on Seyfert 1 galaxies.
Most of these sources were observed nearly weekly whenever they
were accessible, making these and NGC 5548 (Peterson et al.\ 1994)
especially well-suited to the study of long-term 
continuum and emission variability in non-blazar AGNs.
For the best-sampled galaxies in our sample, we have over 100
homogeneous spectra.

Of the nine sources presented in this paper, we were able to 
measure \Hbeta\ response times, or lags, for eight of them. The
lags range from a little more than two weeks to more than
two months, as summarized in Table 7. In many cases, the lags
are measurable with data from individual observing seasons. These
sometimes show variations from year to year, and this probably
indicates that the BLR is physically thick, i.e., the outer
radius of the BLR is much greater than the inner radius.
We have combined the measured BLR response times
with measured widths of the \Hbeta\ rms profiles to obtain virial
mass estimates for the central source. These are all in
the range $10^{7-8}$\,\Msun.

We are grateful for support of this program by the National
Science Foundation under grants AST--9420080, and its predecessors,
AST--8702691, AST--8915258, and AST--9117086. This extensive
program was made possible through the kind cooperation of our
colleagues at Ohio State University and Lowell Observatory.
We thank the current and past Directors of Lowell Observatory,
R.L.\ Millis and J.S.\ Gallagher, for their support of this
project for so many years. Maintaining the CCDS for weekly
observations for nearly a decade has been successful on
account of the capable work of B.\ Atwood, P.L.\ Byard, K.\ Duemmel,
A.A.\ Henden, J.A.\ Mason, T.P.\ O'Brien, and R.J.\ Truax at Ohio State and 
R.\ Nye and R.\ Oliver at Lowell Observatory, and we gratefully
acknowledge their contributions to this program.
We thank several Ohio State students for
their participation in this project: B.\ Ali, 
T.M.\ Kassebaum, K.T.\ Korista, N.J.\ Lame,
P.A.\ Popowski, S.M.\ Smith, and R.J.\ White.
We also thank  M.\ Dietrich for helpful comments and corrections.



\clearpage

\begin{figure}
\caption{The average (left column) and rms (right column) spectra of
3C 120, Akn 120, and Mrk 79, based on the spectra reported here.
The wavelength scale has been divided by $(1+z)$ to put each spectrum in
the rest frame of the source. 
The strong broad line in the center is \Hbeta\,$\lambda4861$, and the
narrow lines to the right are [\oiii]\,$\lambda\lambda4959$, 5007.
In the left column, the horizontal line 
underneath \Hbeta\ shows the integration range
for the line flux, and the bracketed region to the right shows the
continuum region measured (as in Table 3).
The broad \heii\,$\lambda4686$ line
is visible in some of the mean spectra, although
it is often badly blended with \feii\ lines.
The rms spectra highlight the variable parts of the spectrum. In most cases,
the \heii\,$\lambda4686$ line is much more prominent in the rms
spectrum than in the mean spectrum. The rms spectra also enhance the
contrast of variable features in the line profiles, but also amplify
residuals in the [\oiii]\,$\lambda\lambda4959$, 5007, which are used
for flux calibration of these data.}
\end{figure}

\begin{figure}
\caption{The average (left column) and rms (right column) spectra of
Mrk 110, Mrk 335, and Mrk 509, based on the spectra reported here.
The spectra are plotted as described in Fig.\ 1.
The Mrk 509 mean and rms spectra are based on the 52 new spectra 
reported here, plus 95 spectra from the same instrument 
previously included in the presentation of Carone et al.\ (1996).}
\end{figure}

\begin{figure}
\caption{The average (left column) and rms (right column) spectra of
Mrk 590, Mrk 704, and Mrk 817, based on the spectra reported here.
The spectra are plotted as described in Fig.\ 1.}
\end{figure}

\begin{figure}
\caption{The light curves for 3C\,120. Upper panel: the continuum, centered
at 5114\,\AA\ in the rest frame of the source, in units of
$10^{-15}$\,\contunits.
Lower panel: the \Hbeta\ emission line, in units of 
$10^{-13}$\,\lineunits.}
\end{figure}

\begin{figure}
\caption{The light curves for Akn\,120. Upper panel: the continuum, centered
at 5099\,\AA\ in the rest frame of the source, in units of
$10^{-15}$\,\contunits.
Lower panel: the \Hbeta\ emission line, in units of 
$10^{-13}$\,\lineunits.}
\end{figure}

\begin{figure}
\caption{The light curves for Mrk 79. Upper panel: the continuum, centered
at 5090\,\AA\ in the rest frame of the source, in units of
$10^{-15}$\,\contunits.
Lower panel: the \Hbeta\ emission line, in units of 
$10^{-13}$\,\lineunits.}
\end{figure}

\begin{figure}
\caption{The light curves for Mrk 110. Upper panel: the continuum, centered
at 5111\,\AA\ in the rest frame of the source, in units of
$10^{-15}$\,\contunits.
Lower panel: the \Hbeta\ emission line, in units of 
$10^{-13}$\,\lineunits.}
\end{figure}

\begin{figure}
\caption{The light curves for Mrk 335. Upper panel: the continuum, centered
at 5100\,\AA\ in the rest frame of the source, in units of
$10^{-15}$\,\contunits.
Lower panel: the \Hbeta\ emission line, in units of 
$10^{-13}$\,\lineunits.}
\end{figure}

\begin{figure}
\caption{The light curves for Mrk 509. Upper panel: the continuum, centered
at 5126\,\AA\ in the rest frame of the source, in units of
$10^{-15}$\,\contunits.
Lower panel: the \Hbeta\ emission line, in units of 
$10^{-13}$\,\lineunits.}
\end{figure}

\begin{figure}
\caption{The light curves for Mrk 590. Upper panel: the continuum, centered
at 5117\,\AA\ in the rest frame of the source, in units of
$10^{-15}$\,\contunits.
Lower panel: the \Hbeta\ emission line, in units of 
$10^{-13}$\,\lineunits.}
\end{figure}

\begin{figure}
\caption{The light curves for Mrk 704. Upper panel: the continuum, centered
at 5095\,\AA\ in the rest frame of the source, in units of
$10^{-15}$\,\contunits.
Lower panel: the \Hbeta\ emission line, in units of 
$10^{-13}$\,\lineunits.}
\end{figure}

\begin{figure}
\caption{The light curves for Mrk 817. Upper panel: the continuum, centered
at 5046\,\AA\ in the rest frame of the source, in units of
$10^{-15}$\,\contunits.
Lower panel: the \Hbeta\ emission line, in units of 
$10^{-13}$\,\lineunits.}
\end{figure}

\begin{figure}
\caption{Cross-correlation functions based on the entire light
curves shown in Figs.\ 4--10 and 12. The solid line represents
the interpolated cross-correlation function (ICCF) and the
individual vertical lines show values of the discrete correlation function
(DCF) and associated errors. The two methods are in good agreement
on the time scales of interest, i.e., less than 100 days, which
is where the CCFs peak in each case. Deviations at larger lags
arise on account of the gaps between observing seasons.
The lags derived from the ICCFs are shown in Table 6.
Of all the sources reported here, only the poorly sampled light
curve of Mrk 704 failed to yield a statistically significant lag.}
\end{figure}

\begin{figure}
\caption{Cross-correlation functions based on the subsets
that yield the smallest formal errors. Only the interpolation
cross-correlation function is shown since these are based on
too few points for the DCF to perform well (see White \& Peterson 1994).
Many of these CCFs
show weak sharp features at $\tau = 0$, indicative of
weak correlated (positive peaks) or anticorrelated (negative peaks)
errors.}
\end{figure}

\begin{figure}
\caption{The combined light curves for Akn\,120,
including IDS data from Peterson, Korista, \& Wagner (1989),
and CCD data, as in Fig.\ 5. The uncertainties in the CCD data
are about four times smaller than the uncertainties in the IDS data.
Upper panel: the continuum, centered
at $\sim5100$\,\AA\ in the rest frame of the source, in units of
$10^{-15}$\,\contunits.
Lower panel: the \Hbeta\ emission line, in units of 
$10^{-13}$\,\lineunits.}
\end{figure}

\begin{figure}
\caption{Cross-correlation functions for Akn 120, based
on IDS data from Peterson, Korista, \& Wagner (1989) as
shown in Fig.\ 14 (data points prior to JD2447400).
The solid line represents the interpolated cross-correlation
function (ICCF), and the points with associated error bars
are discrete correlation function (DCF) values. This should
be compared with Fig.\ $5b$ of Edelson \& Krolik (1988).
}
\end{figure}

\clearpage


\begin{deluxetable}{lccccc}
\tablewidth{0pt}
\tablecaption{List of Sources}
\label{tab:sourcetab}
\tablehead{
\colhead{ } &
\multicolumn{2}{c}{Position (1950.0)} &
\colhead{ } &
\colhead{$A_B$} &
\colhead{$L_{\lambda}$\,(5100\,\AA)} \nl
\colhead{Source } & 
\colhead{$\alpha$} &
\colhead{$\delta$} &
\colhead{$z$} &
\colhead{(mag)} &
\colhead{(ergs s$^{-1}$ \AA$^{-1}$)} \nl
\colhead{(1)} & 
\colhead{(2)} & 
\colhead{(3)} & 
\colhead{(4)} & 
\colhead{(5)} &
\colhead{(6)}
} 
\startdata
Mrk\,335 	& $00^{\rm h}\,03^{\rm m}\,45^{\rm s}\!.2$ 
		& $+19^{\rm o}\,55^{'}\,29^{''}$
		& $0.026$ & $0.10$ & $6.6\times10^{39}$ \nl
Mrk\,590 	& $02^{\rm h}\,12^{\rm m}\,00^{\rm s}\!.4$ 
		& $-00^{\rm o}\,59^{'}\,58^{''}$
		& $0.026$ & $0.05$ & $5.4\times10^{39}$ \nl
3C\,120 	& $04^{\rm h}\,30^{\rm m}\,31^{\rm s}\!.6$  
		& $+05^{\rm o}\,15^{'}\,00^{''}$
		& $0.033$ & $0.57$ & $7.6\times10^{39}$ \nl
Akn\,120 	& $05^{\rm h}\,13^{\rm m}\,37^{\rm s}\!.9$ 
		& $-00^{\rm o}\,12^{'}\,15^{''}$
		& $0.033$ & $0.40$ &$1.5\times10^{40}$ \nl
Mrk\,79  	& $07^{\rm h}\,38^{\rm m}\,47^{\rm s}\!.3$ 
		& $+49^{\rm o}\,55^{'}\,41^{''}$
		& $0.022$ & $0.23$ & $4.5\times10^{39}$ \nl
Mrk\,704 	& $09^{\rm h}\,15^{\rm m}\,39^{\rm s}\!.4$ 
		& $+16^{\rm o}\,30^{'}\,59^{''}$
		& $0.030$ & $0.05$ & $4.9\times10^{39}$ \nl
Mrk\,110 	& $09^{\rm h}\,21^{\rm m}\,44^{\rm s}\!.4$ 
		& $+52^{\rm o}\,30^{'}\,08^{''}$
		& $0.035$ & $0$ & $4.0\times10^{39}$ \nl
Mrk\,817 	& $14^{\rm h}\,34^{\rm m}\,57^{\rm s}\!.9$ 
		& $+59^{\rm o}\,00^{'}\,39^{''}$
		& $0.031$ & $0$ & $5.6\times10^{39}$ \nl
Mrk\,509 	& $20^{\rm h}\,41^{\rm m}\,26^{\rm s}\!.3$ 
		& $-10^{\rm o}\,54^{'}\,18^{''}$
		& $0.034$ & $0.18$ & $1.6\times10^{40}$ \nl

\enddata
\end{deluxetable}


\begin{deluxetable}{lcc}
\tablewidth{0pt}
\tablecaption{Absolute [O\,{\sc iii}]\,$\lambda5007$ Fluxes}
\label{tab:abscal}
\tablehead{
\colhead{ } &
\colhead{Number of} &
\colhead{Flux} \nl
\colhead{Source} & 
\colhead{Observations} & 
\colhead{($10^{-13}$\,ergs s$^{-1}$ cm$^{-2}$)} \nl
\colhead{(1)} & 
\colhead{(2)} & 
\colhead{(3)}
} 
\startdata
3C\,120 	& 12	& $3.02 \pm 0.16$  \nl
Akn 120		& 28    & $0.91 \pm 0.04$  \nl
Mrk 79		& 28    & $3.16 \pm 0.19$  \nl
Mrk 110		& 26    & $2.26 \pm 0.14$  \nl
Mrk 335		& 13    & $2.31 \pm 0.10$  \nl
Mrk 509 	& 13	& $6.79 \pm 0.40$\tablenotemark{a}  \nl
Mrk 590		& 17 	& $1.04 \pm 0.05$  \nl
Mrk 704		& 17	& $1.27 \pm 0.07$  \nl
Mrk 817		& 20	& $1.34 \pm 0.05$  
\enddata
\tablenotetext{a}{From Carone et al.\ (1996).}
\end{deluxetable}

\begin{deluxetable}{lccccc}
\tablewidth{0pt}
\tablecaption{Integration Limits}
\label{tab:tabwave}
\tablehead{
\colhead{ } & 
\multicolumn{3}{c}{Wavelength Range (\AA) } \nl
\colhead{Object} &
\colhead{Continuum} & 
\colhead{H$\beta$} & 
\colhead{Continuum Under H$\beta$} \nl
\colhead{(1)} & 
\colhead{(2)} & 
\colhead{(3)} & 
\colhead{(4)} 
} 
\startdata
3C\,120  & 5270--5295 & 4930--5100 & 4925--5287  \nl
Akn\,120 & 5260--5275 & 4915--5105 & 4905--5267  \nl
Mrk\,79  & 5195--5210 & 4845--5042 & 4840--5202  \nl
Mrk\,110 & 5280--5300 & 4975--5110 & 4965--5290  \nl
Mrk\,335 & 5225--5240 & 4900--5038 & 4895--5232  \nl
Mrk\,509 & 5290--5310 & 4955--5110 & 4950--5300  \nl
Mrk\,590 & 5240--5260 & 4895--5045 & 4892--5250  \nl
Mrk\,704 & 5240--5255 & 4900--5080 & 4895--5252  \nl
Mrk\,817 & 5250--5275 & 4915--5090 & 4905--5262
\enddata
\end{deluxetable}

\begin{deluxetable}{lcccc}
\tablewidth{0pt}
\tablecaption{Sampling Statistics}
\label{tab:sampstats}
\tablehead{
\colhead{} & 
\colhead{No.} & 
\colhead{Span} & 
\colhead{$\langle\Delta T \rangle$} & 
\colhead{$\Delta T_{\rm med}$} \nl
\colhead{Source} &
\colhead{Obs.} &
\colhead{(days)} &
\colhead{(days)} &
\colhead{(days)}  \nl
\colhead{(1)} & 
\colhead{(2)} & 
\colhead{(3)} &
\colhead{(4)} & 
\colhead{(5)}  \nl
} 
\startdata
3C\,120 & 52 & 2206 & 50 & 11  \nl
Akn\,120& 141& 2864 & 20 & 8   \nl
Mrk\,79 &143 & 2829 & 20 & 8   \nl
Mrk\,110&95  & 1737 & 18 & 8   \nl
Mrk\,335&123 & 2600 & 21 & 7   \nl
Mrk\,509&52  & 728  & 18 & 7\tablenotemark{a}   \nl
Mrk\,509&194 & 2959 & 15 & 7\tablenotemark{b}   \nl
Mrk\,590&102 & 2551 & 25 & 8   \nl
Mrk\,704&29  & 1457 & 52 & 20  \nl
Mrk\,817&81  & 1493 & 19 & 7   
\enddata
\tablenotetext{a}{New data only.}
\tablenotetext{b}{New data and data from Carone et al.\ (1996).}
\end{deluxetable}

\begin{deluxetable}{lcccccc}
\tablewidth{0pt}
\tablecaption{Variability Statistics}
\label{tab:varstats}
\tablehead{
\colhead{} & 
\multicolumn{3}{c}{Continuum} &
\multicolumn{3}{c}{H$\beta$} \nl
\colhead{Source} &
\colhead{$\langle F \rangle$\tablenotemark{a}} & 
\colhead{$F_{var}$} & 
\colhead{$R_{max}$} & 
\colhead{$\langle F \rangle$\tablenotemark{b}} & 
\colhead{$F_{var}$} & 
\colhead{$R_{max}$} \nl
\colhead{(1)} & 
\colhead{(2)} & 
\colhead{(3)} &
\colhead{(4)} & 
\colhead{(5)} & 
\colhead{(6)} & 
\colhead{(7)} 
} 
\startdata
3C\,120 	& \mone$4.30\pm0.77$ & 0.178 & $2.34\pm0.07$ 
		& \mone$3.78\pm0.37$ & 0.095 & $1.65\pm0.05$ \nl
Akn\,120	& \mone$9.31\pm1.74$ & 0.186 & $1.93\pm0.06$ 
		& \mone$8.74\pm1.62$ & 0.184 & $2.03\pm0.06$ \nl
Mrk\,79		& \mone$7.31\pm0.96$ & 0.130 & $1.92\pm0.05$ 
		& \mone$5.56\pm0.37$ & 0.062 & $1.44\pm0.04$ \nl
Mrk\,110	& \mone$3.05\pm1.02$ & 0.334 & $3.64\pm0.11$ 
		& \mone$3.40\pm0.85$ & 0.249 & $2.74\pm0.16$ \nl
Mrk\,335	& \mone$8.44\pm0.78$ & 0.090 & $1.55\pm0.04$
		& \mone$8.21\pm0.52$ & 0.060 & $1.34\pm0.04$ \nl
Mrk\,509\tablenotemark{c}	
		& \mone$9.65\pm0.85$ & 0.085 & $1.39\pm0.04$
		& $10.89\pm0.65$& 0.056 & $1.26\pm0.04$ \nl
Mrk\,509\tablenotemark{d}	
		& $10.92\pm1.87$& 0.168 & $2.15\pm0.06$
		& $11.94\pm1.29$& 0.105 & $1.81\pm0.06$ \nl
Mrk\,590	& \mone$7.18\pm1.35$ & 0.187 & $2.18\pm0.06$
		& \mone$3.66\pm1.05$ & 0.286 & $4.37\pm0.12$ \nl
Mrk\,704	& \mone$4.84\pm0.46$ & 0.092 & $1.37\pm0.04$
		& \mone$3.39\pm0.30$ & 0.087 & $1.36\pm0.04$ \nl
Mrk\,817        & \mone$5.36\pm0.78$ & 0.144 & $1.91\pm0.05$ 
		& \mone$3.96\pm0.66$ & 0.166 & $1.82\pm0.05$
\enddata
\tablenotetext{a}{Units $10^{-15}$\,ergs s$^{-1}$\,cm$^{-2}$\,\AA$^{-1}$.}
\tablenotetext{b}{Units $10^{-13}$\,ergs s$^{-1}$\,cm$^{-2}$.}
\tablenotetext{c}{New data only.}
\tablenotetext{d}{New data and data from Carone et al.\ (1996).}
\end{deluxetable}

\begin{deluxetable}{llllccc}
\tablewidth{0pt}
\tablecaption{Cross-Correlation Results}
\label{tab:ccfs}
\tablehead{
\colhead{} 		&
\colhead{} 		&
\colhead{$\tau_{cent}$} &
\colhead{$\tau_{peak}$} &
\colhead{ }		&
\colhead{FWHM}		&
\colhead{Number}		\nl
\colhead{Source} &
\colhead{Subset} &
\colhead{(days)} 	&
\colhead{(days)} 	&
\colhead{$r_{max}$}	&
\colhead{(days)}	&
\colhead{Obs.}		\nl
\colhead{(1)} & 
\colhead{(2)} & 
\colhead{(3)} &
\colhead{(4)} & 
\colhead{(5)} & 
\colhead{(6)} & 
\colhead{(7)} 
} 
\startdata
3C\,120         & All data
		& $43.8^{+27.7}_{-20.3}$ & $34^{+10}_{-12}$ 
		& 0.674 & 406  & 52 \nl
Akn\,120 	& All data
		& $60.2^{+31.1}_{-13.2}$ & $52^{+2}_{-14}$  
 		& 0.934 & 874  & 141 \nl
Akn\,120& 48149--48345 	
		& $49.5^{+12.4}_{-14.6}$ & $54^{+11}_{-16}$  
 		& 0.857 & 71 & 20  \nl
Akn\,120& 48870--49090 	
		& $31.6^{+14.1}_{-12.6} $& $32^{+16}_{-14}$  
 		& 0.767 & 45 & 23  \nl
Akn\,120& 49981--50176	
		& $38.6^{+5.3}_{-6.5} $ & $29^{+24}_{-1}$  
 		& 0.944 & 71 & 20  \nl
Mrk\,79 & All data
		& $25.7^{+11.7}_{-4.8} $& $31^{+4}_{-6}$  
 		& 0.664 & 353 & 143  \nl
Mrk\,79 & 47838--48044
		& $10.4^{+10.5}_{-11.4} $& $14^{+9}_{-13}$  
 		& 0.622 & 44 & 20  \nl
Mrk\,79 & 48193--48393
		& $18.1^{+4.9}_{-8.6} $& $28^{+1}_{-21}$  
 		& 0.854 & 63 & 19  \nl
Mrk\,79 & 48905--49135
		& $16.1^{+16.0}_{-7.0} $& $15^{+19}_{-6}$  
 		& 0.702 & 61 & 23  \nl
Mrk\,79 & 49996--50220
		& $41.6^{+6.2}_{-28.9} $& $44^{+5}_{-33}$  
 		& 0.710 & 59 & 24  \nl
Mrk\,110& All data
		& $31.6^{+9.0}_{-7.3} $& $25^{+5}_{-5}$   
 		& 0.896 & 300 & 95  \nl
Mrk\,110& 48954--49149
		& $27.5^{+4.8}_{-23.9}$ & $27^{+6}_{-51}$   
 		& 0.746 & 24 & 21  \nl
Mrk\,110& 49752--49875
		& $19.5^{+6.5}_{-6.8} $& $20^{+14}_{-6}$   
 		& 0.718 & 37 & 14  \nl
Mrk\,110& 50011--50262
		& $50.7^{+0.8}_{-27.5} $& $24^{+6}_{-3}$   
 		& 0.969 & 213& 28  \nl
Mrk\,335& All data
		& $16.8^{+5.2}_{-3.3} $& $12^{+7}_{-2}$   
 		& 0.660 & 112 & 123  \nl
Mrk\,335& 49156--49338
		& $15.6^{+6.9}_{-3.4} $& $18^{+5}_{-9}$   
 		& 0.869 & 38 & 24  \nl
Mrk\,335& 49889--50118
		& $12.5^{+7.1}_{-5.1} $& $15^{+5}_{-9}$   
 		& 0.791 & 30 & 25  \nl
Mrk\,509& All data\tablenotemark{a}
		& $79.3^{+6.5}_{-6.2}   $& $86^{+1}_{-20}$ 
		& 0.858 & 216 & 194 \nl
Mrk\,590& All data
		& $22.5^{+17.8}_{-18.1} $& $25^{+2}_{-2}$   
		& 0.936 & 420 & 102 \nl
Mrk\,590& 48090--48323
		& $20.5^{+4.5}_{-3.0} $& $21^{+5}_{-4}$   
		& 0.716 & 24  & 24  \nl
Mrk\,590& 48848--49048
		& $15.7^{+9.0}_{-11.7} $& $17^{+8}_{-11}$   
		& 0.959 & 149 & 17  \nl
Mrk\,590& 49183--49338
		& $28.7^{+7.2}_{-4.5} $& $34^{+2}_{-12}$   
		& 0.884 & 40  & 16  \nl
Mrk\,590& 49958--50122
		& $28.5^{+5.0}_{-3.5} $& $27^{+10}_{-2}$   
   		& 0.937 & 43  & 17  \nl
Mrk\,817& 49404 onward
		& $27.0^{+10.1}_{-3.9} $& $33^{+3}_{-15}$   
		& 0.702 & 84  & 47  \nl
Mrk\,817& 49000--49211
		& $19.5^{+4.5}_{-4.1} $& $21^{+2}_{-6}$    
		& 0.827 & 95  & 25  \nl
Mrk\,817& 49404--49527
		& $15.5^{+4.3}_{-3.5} $& $17^{+4}_{-5}$    
		& 0.903 & 35  & 17  \nl
Mrk\,817& 49752--49923
		& $36.3^{+8.0}_{-8.7} $& $35^{+12}_{-7}$  
		& 0.876 & 50  & 19  \nl

\enddata
\tablenotetext{a}{New data and data from Carone et al.\ (1996).}
\end{deluxetable}

\begin{deluxetable}{lccc}
\tablewidth{0pt}
\tablecaption{Broad-Line Region Size and Virial Mass Estimates}
\label{tab:ccfs}
\tablehead{
\colhead{} 		&
\colhead{$\tau_{cent}$} &
\colhead{\vFWHM(rms)}   &
\colhead{Mass}		\nl
\colhead{Source} &
\colhead{(days)} 	&
\colhead{(\kms)} 	&
\colhead{($10^7$\,\Msun)}	\nl
\colhead{(1)} & 
\colhead{(2)} & 
\colhead{(3)} &
\colhead{(4)} 
} 
\startdata
3C\,120         & $43.8^{+27.7}_{-20.3}$ & 2300 & \mone3.4 \nl
Akn\,120	& $38.6^{+5.3}_{-6.5} $  & 5500 & 17.0 \nl
Mrk\,79 	& $18.1^{+4.9}_{-8.6} $  & 6200 & 10.1 \nl
Mrk\,110	& $19.5^{+6.5}_{-6.8} $  & 2500 & \mone1.8 \nl
Mrk\,335	& $16.8^{+5.2}_{-3.3} $  & 1800 & \mone0.8 \nl
Mrk\,509	& $79.3^{+6.5}_{-6.2} $  & 2800 & \mone9.0 \nl
Mrk\,590	& $20.5^{+4.5}_{-3.0} $  & 2300 & \mone1.6 \nl
Mrk\,817	& $15.5^{+4.3}_{-3.5} $  & 4100 & \mone3.8 \nl
\enddata
\end{deluxetable}


\clearpage

\begin{references}
\reference{}Alloin, D., Clavel, J., Peterson, B.M., Reichert, G.A., \& 
Stirpe, G.M. 1994, in Frontiers of Space and Ground-Based Astronomy, ed.\ 
W. Wamsteker, M.S. Longair, \& Y. Kondo  (Dordrecht: Kluwer), p.\ 423
\reference{}Andrillat, Y., \& Collin-Souffrin, S. 1968,
Ap.\ Letters, 1, 111
\reference{}Blandford, R.D., \& McKee, C.F. 1982, ApJ, 255, 419
\reference{}Byard, P.L., Foltz, C.B., Jenkner, H., \& Peterson, B.M. 1981,
PASP, 93, 147
\reference{}Carone, T.E., et al. 1996, ApJ, 471, 737
\reference{}Clavel, J., et al. 1991, ApJ, 366, 64
\reference{}Collier, S.J., et al. 1998, ApJS, in press
\reference{}Dietrich, M., et al. 1998, ApJS, in press
\reference{}Edelson, R.A., \& Krolik, J.H. 1988, ApJ, 333, 646
\reference{}Ferland, G.J., Korista, K.T., \& Peterson, B.M. 1990,
ApJ, 363, L21
\reference{}Fitch, W., Pacholczyk, A.G., \& Weymann, R.J. 1967,
ApJ, 150, L67
\reference{}French, H.B., \& Miller, J.S. 1980, PASP, 92, 753
\reference{}Gaskell, C.M., \& Peterson, B.M. 1987, ApJS, 65, 1
\reference{}Gaskell, C.M., \& Sparke, L.S. 1986, ApJ,305, 175
\reference{}Kaspi, S., Smith, P.S., Maoz, D., Netzer, H., \& Jannuzi, B.T.,
     1996a, ApJ, 471, L75 
\reference{}Kaspi, S., et al. 1996b, ApJ, 470, 336
\reference{}Kassebaum, T.M., Peterson, B.M., Wanders, I., Pogge, R.W.,
Bertram, R., \& Wagner, R.M. 1997, ApJ, 475, 106
\reference{}Korista, K.T., et al. 1995, ApJS, 97, 285
\reference{}Maoz, D., \& Netzer, H. 1989, MNRAS, 236, 21 
\reference{}Netzer, H. 1990, in Active Galactic Nuclei, 
R.D.\ Blandford, H.\ Netzer, and L.\ Woltjer (Berlin: Springer-Verlag),
p.\ 137
\reference{}Netzer, H., \& Maoz, D. 1990, ApJ, 365, L5
\reference{}Netzer, H., Maoz, D., Laor, A., Mendelson, H., Brosch, N.,
Leibowitz, E., Almoznino, E., Beck, S., \& Mazeh, T. 1990, ApJ, 353, 108
\reference{}Netzer, H., \& Peterson, B.M. 1997, in Astronomical Time Series,
ed.\ D. Maoz, A. Sternberg, \& E. Leibowitz (Dordrecht: Kluwer), p.\ 85
\reference{}Oke, J.B., Readhead, A.C.S., \& Sargent, W.L.W. 1980,
PASP, 92, 758
\reference{}Penston, M.V. 1991, in Variability of Galactic Nuclei,
ed.\ H.R.\ Miller \& P.J.\ Wiita (Cambridge: Cambridge Univ.\ Press),
p.\ 343
\reference{}Peterson, B.M. 1988, PASP, 100, 18
\reference{}Peterson, B.M. 1993, PASP, 105, 247
\reference{}Peterson, B.M., et al. 1991, ApJ, 368, 119
\reference{}Peterson, B.M., et al. 1992, ApJ, 392, 470
\reference{}Peterson, B.M., et al. 1994, ApJ,  425, 622
\reference{}Peterson, B.M., Ali, B., Horne, K., Bertram, R.,
Lame, N.J., Pogge, R.W., \& Wagner, R.M. 1993, ApJ, 402, 469
\reference{}Peterson, B.M., Crenshaw, D.M., \& Meyers, K.A. 1985,
ApJ, 298, 283
\reference{}Peterson, B.M., \& Gaskell, C.M. 1986, AJ, 92, 552
\reference{}Peterson, B.M., \& Gaskell, C.M. 1991, ApJ, 368, 152
\reference{}Peterson, B.M., Korista, K.T., \&  Wagner, R.M. 1989, AJ,
98. 500
\reference{}Peterson, B.M., Meyers, K.A., Capriotti, E.R., Foltz, C.B.,
Wilkes, B.J., \& Miller, H.R. 1985, ApJ, 292, 164
\reference{}Peterson, B.M., Pogge, R.W., Wanders, I., Smith, S.M.,
\& Romanishin, W. 1995, PASP, 107, 579
\reference{}Peterson, B.M., Wanders, I., Horne, K., Collier, S.,
Alexander, T., \& Kaspi, S. 1998, submitted to PASP
\reference{}Robinson, A. 1994, in Reverberation Mapping of the 
Broad-Line Region in Active Galactic Nuclei, ed.\ P.M.\ Gondhalekar,
K.\ Horne, \&  B.M.\ Peterson, (San Francisco: Astronomical Society
of the Pacific), p.\ 147
\reference{}Rodr\'{\i}guez, P.M., et al. 1997, ApJ, 110, 9
\reference{}Shuder, J.M. 1981, AJ, 86, 1595
\reference{}Stirpe, G.M., et al. 1994, ApJ, 425, 609
\reference{}van Groningen, E., \& Wanders, I. 1992, PASP, 104, 700
\reference{}Wanders, I. 1995, A\&A, 296, 332
\reference{}Wanders, I., Peterson, B.M., Pogge, R.W., DeRobertis, M.M.,
\& van Groningen, E. 1992, A\&A, 266, 72
\reference{}White, R.J., \& Peterson, B.M. 1994, PASP, 106, 879
\end{references}
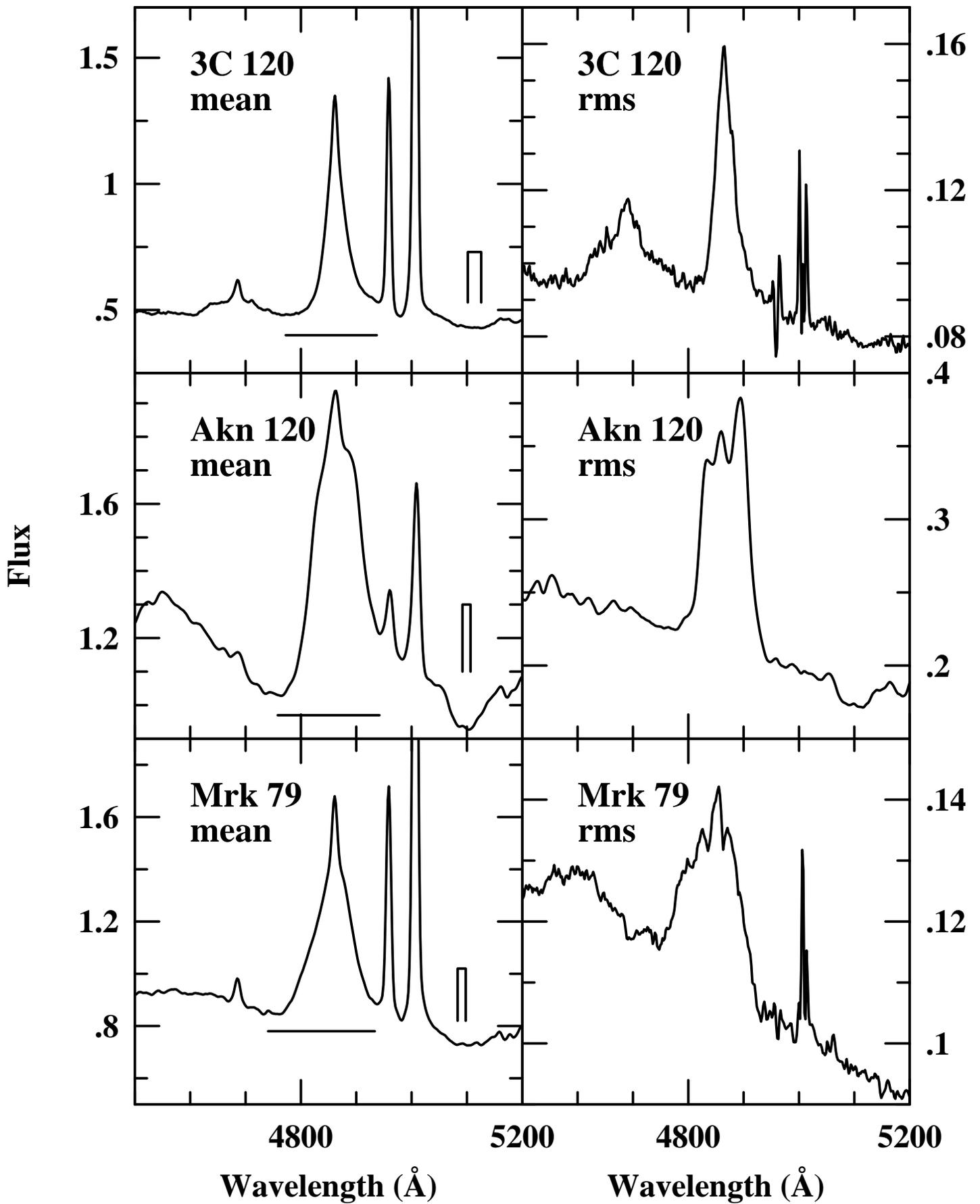
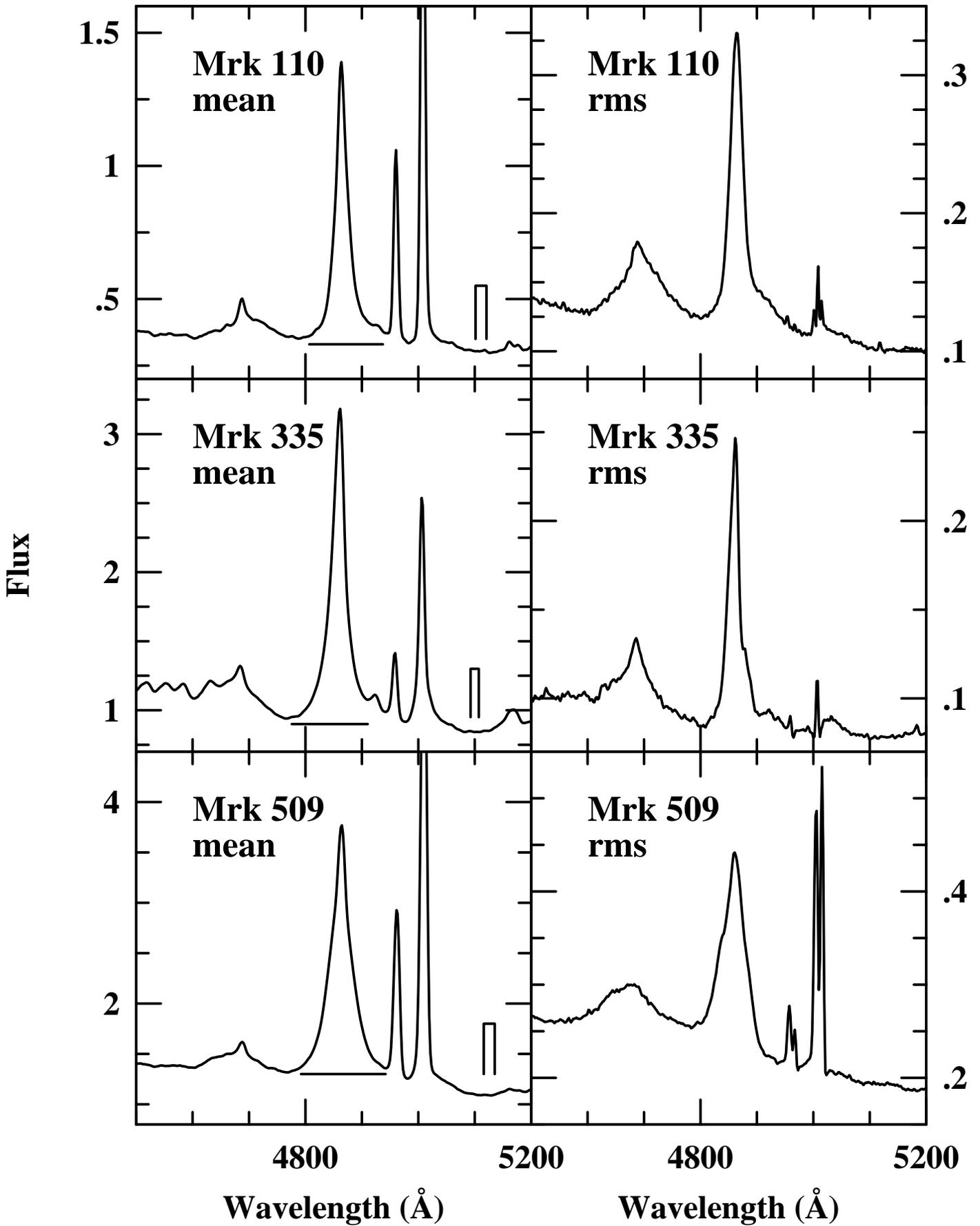
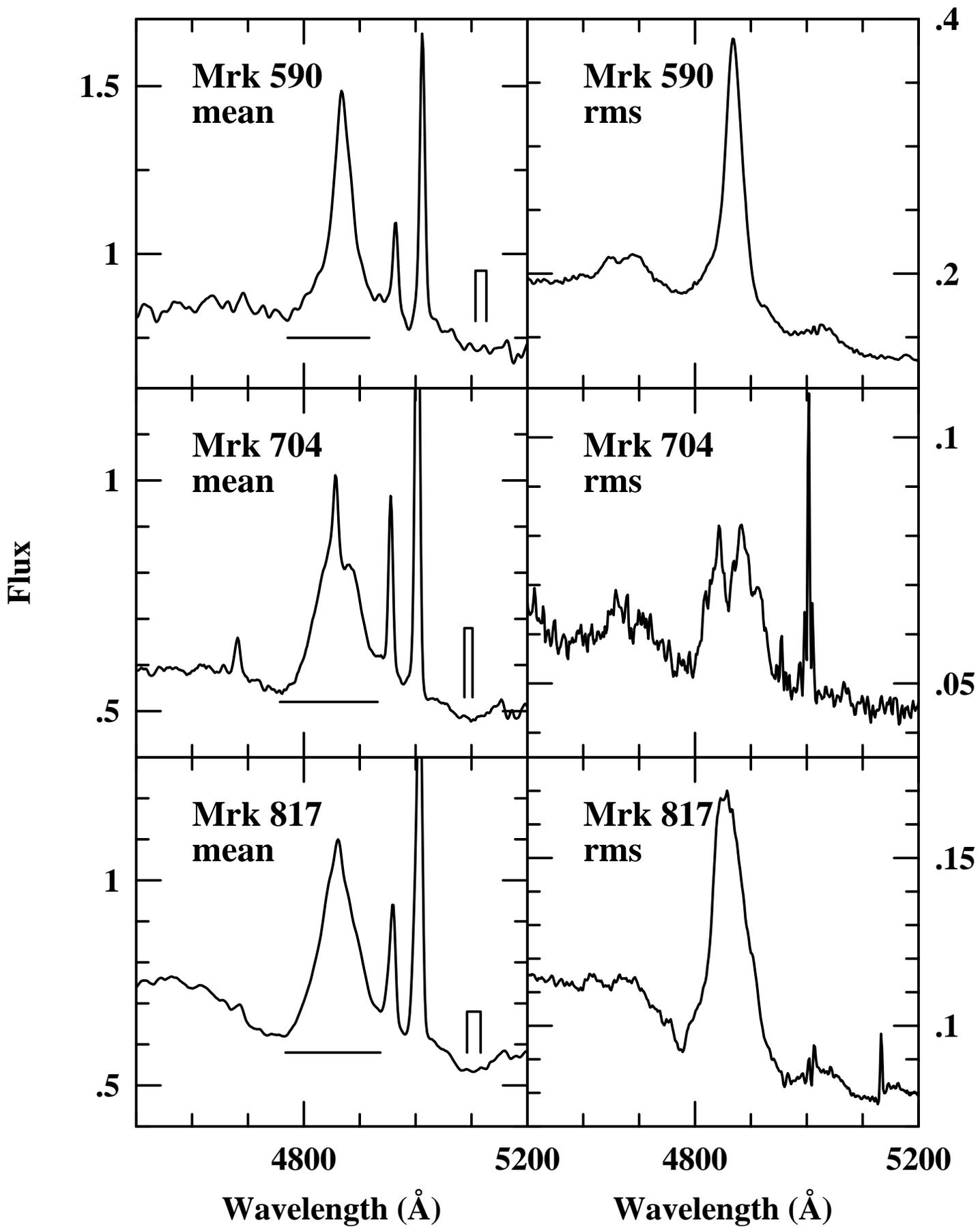
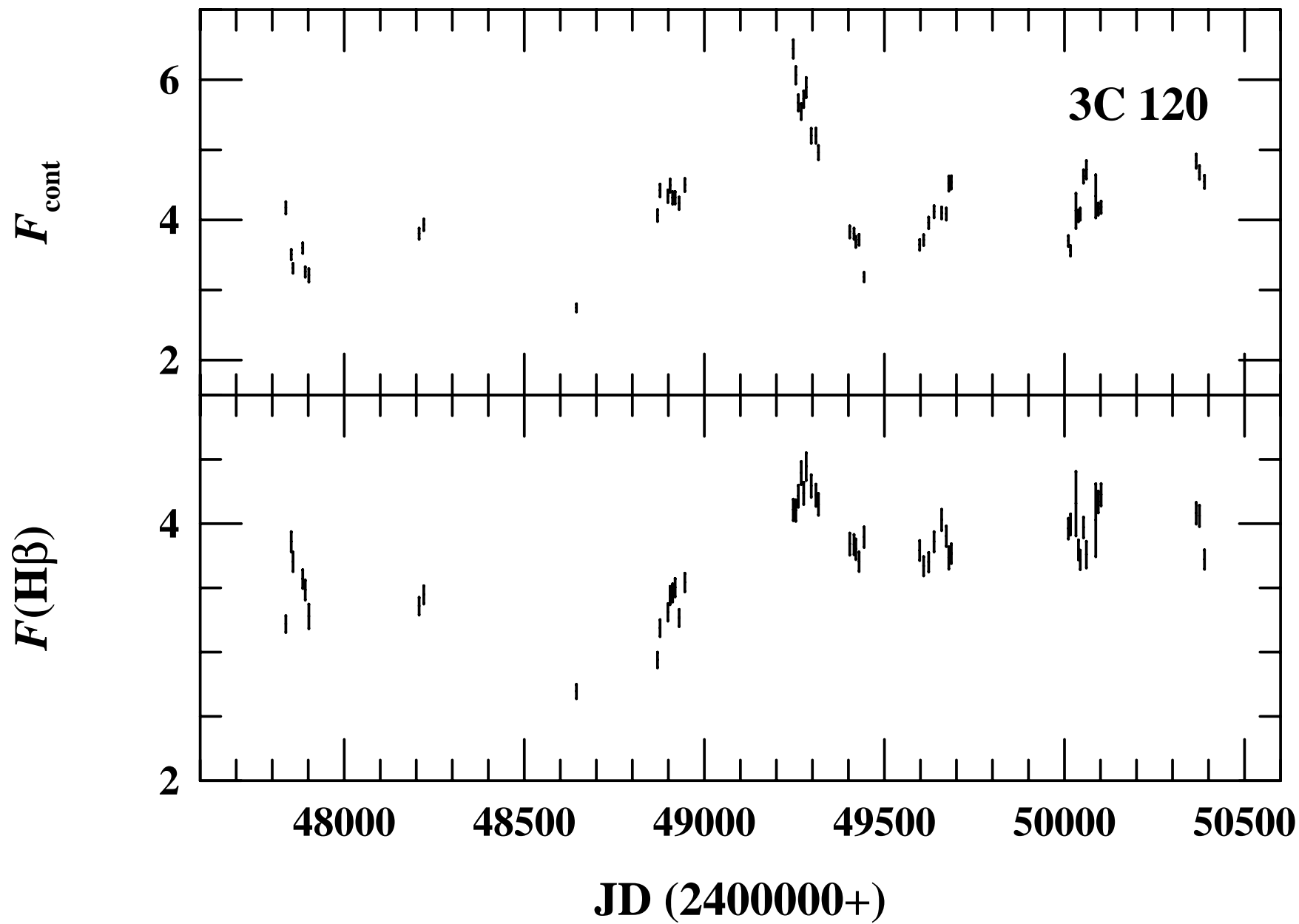
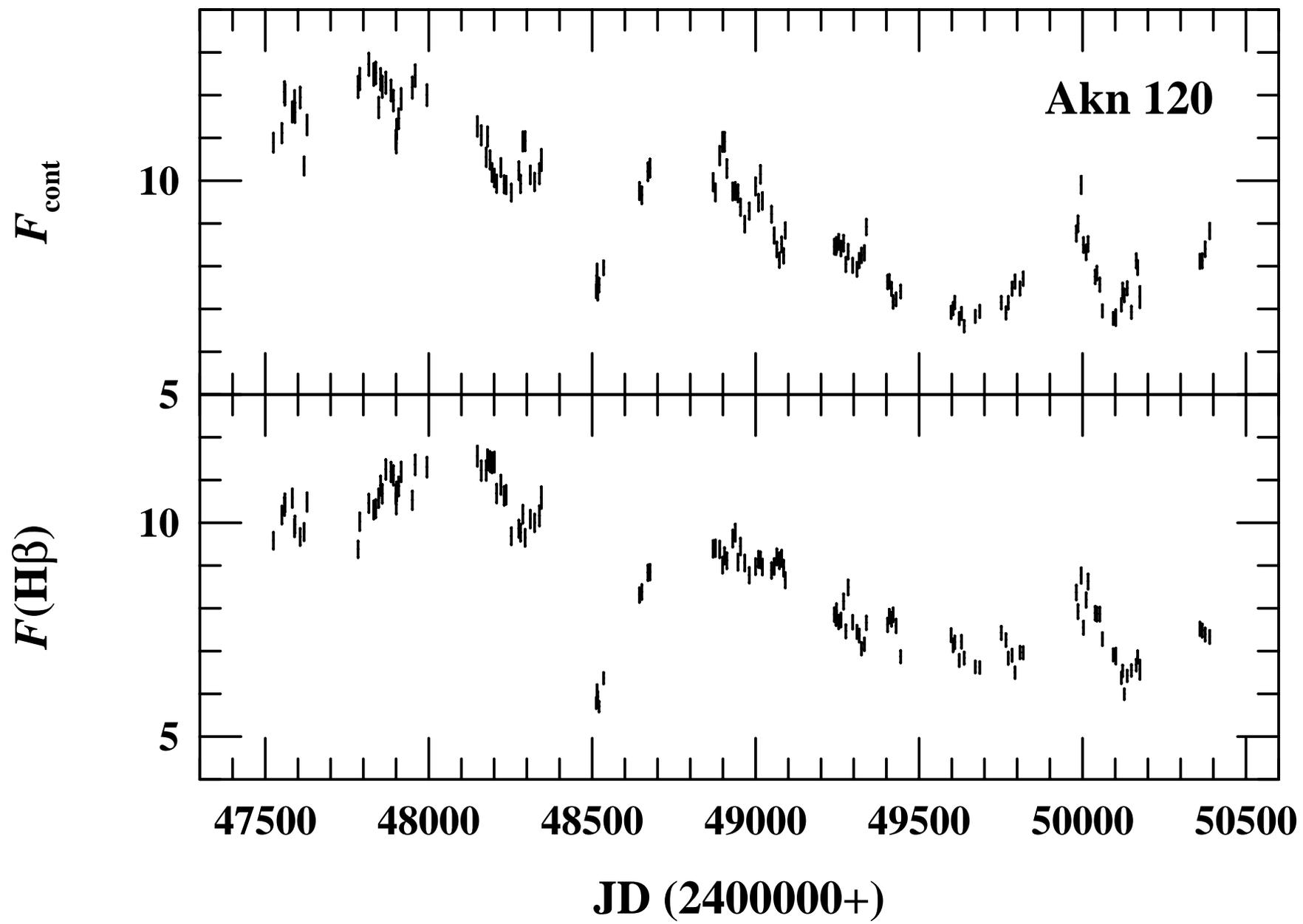
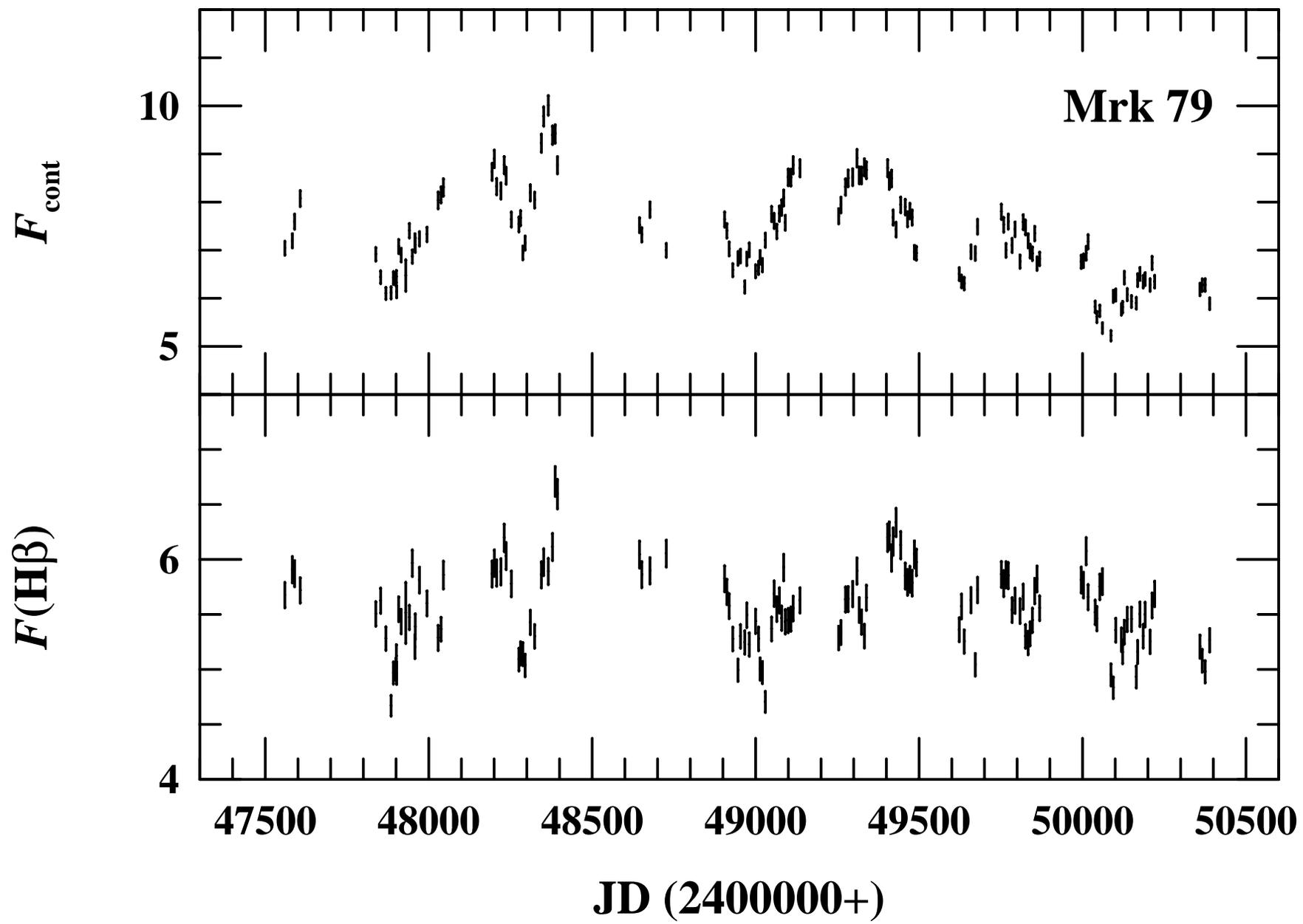
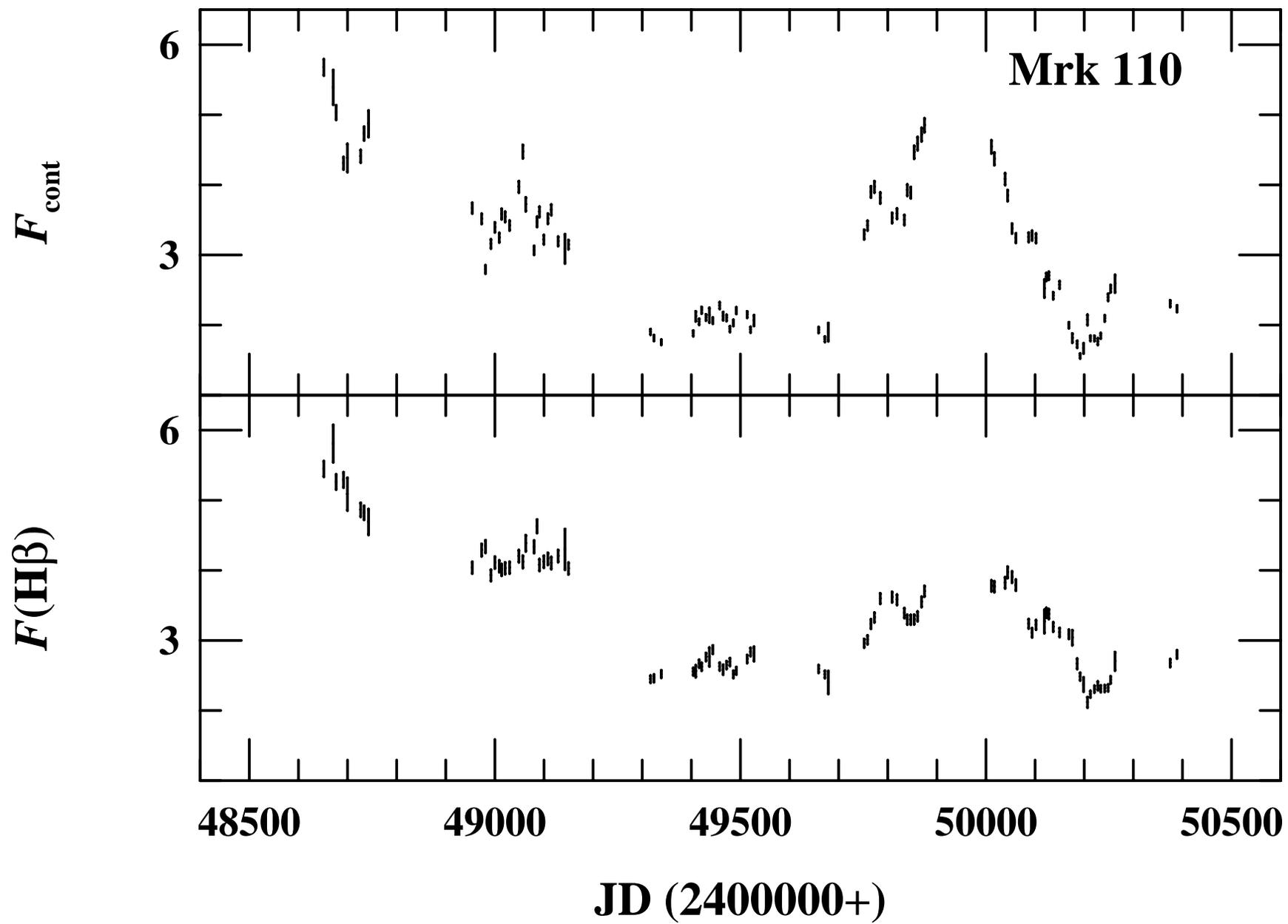
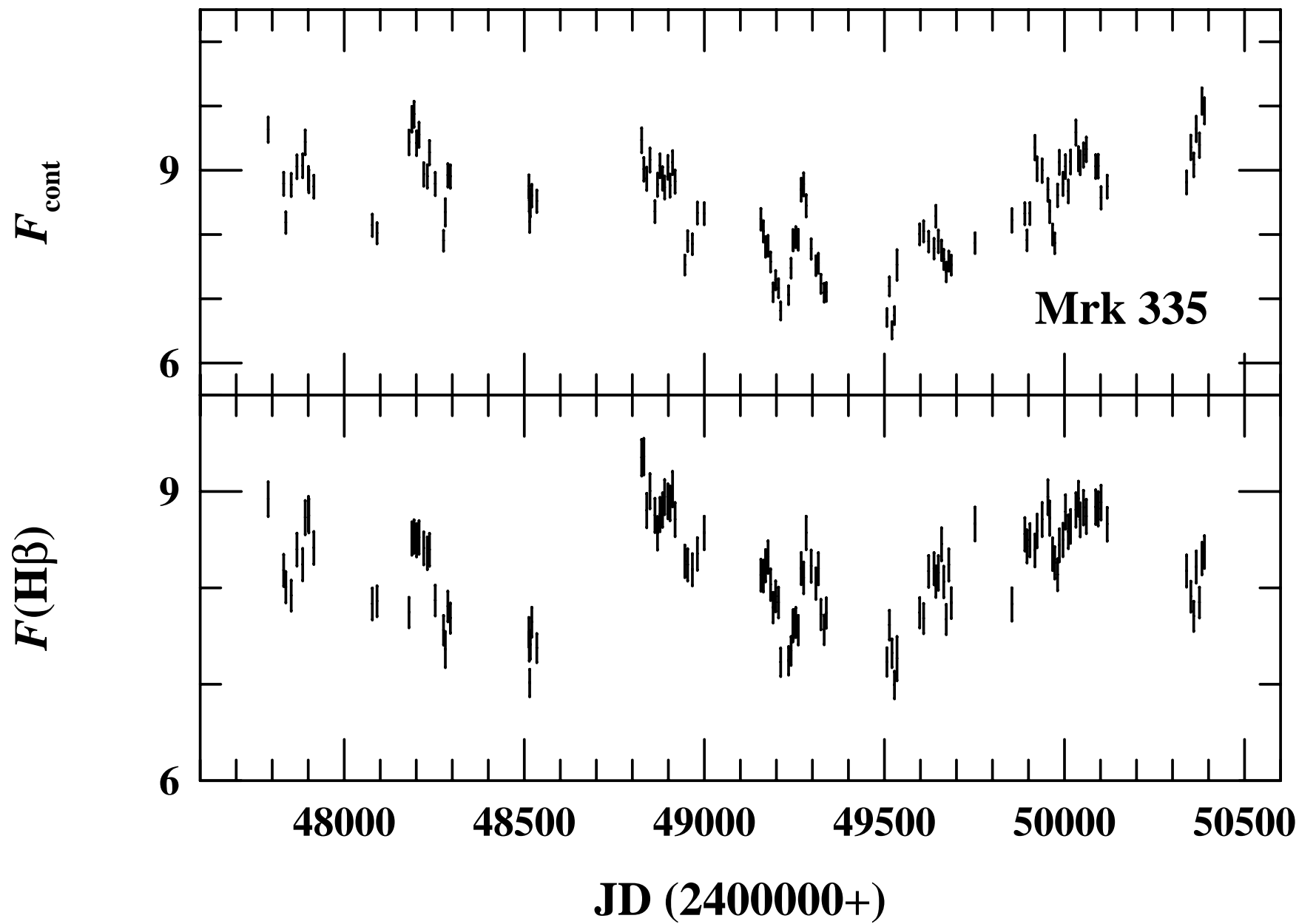
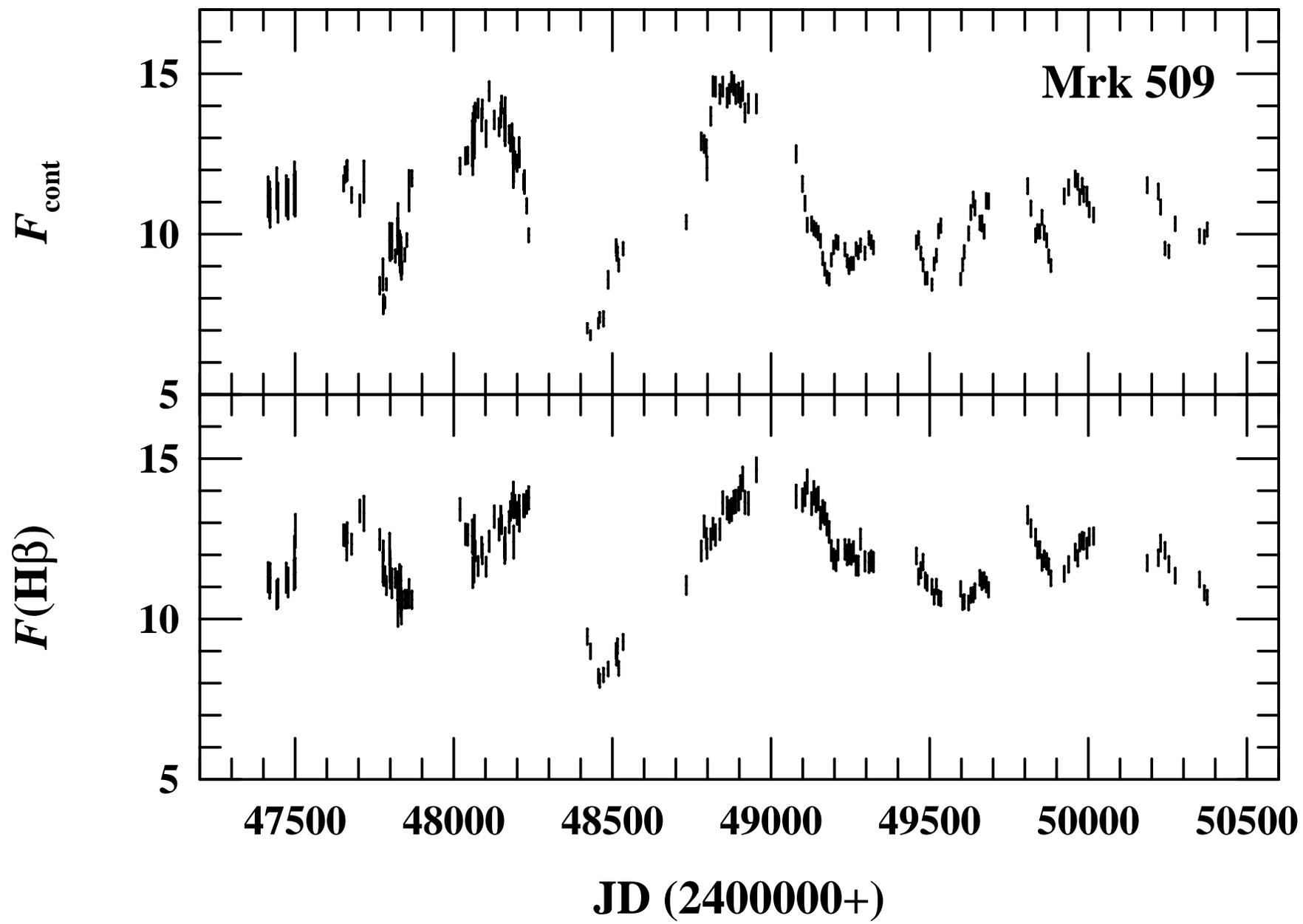
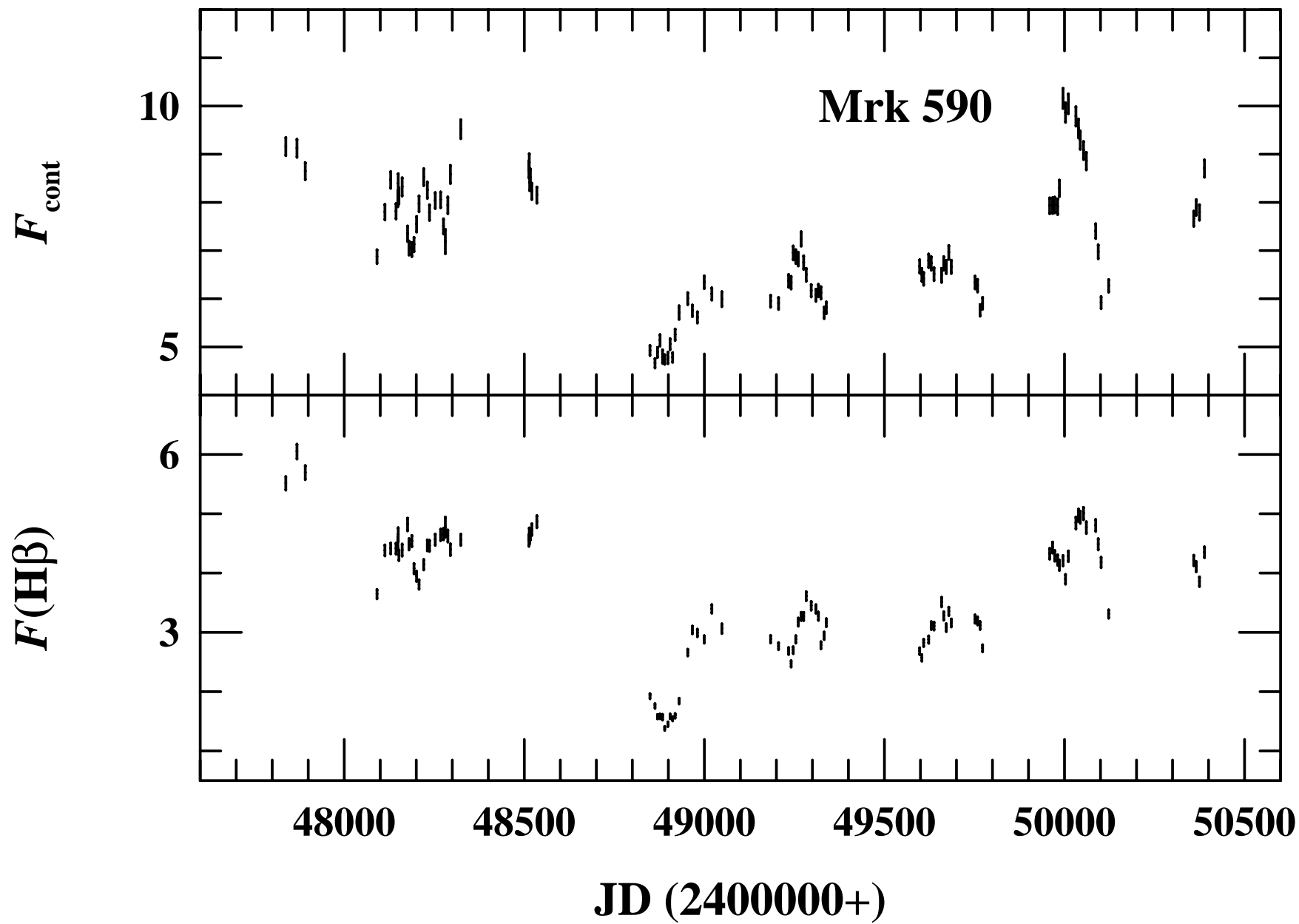
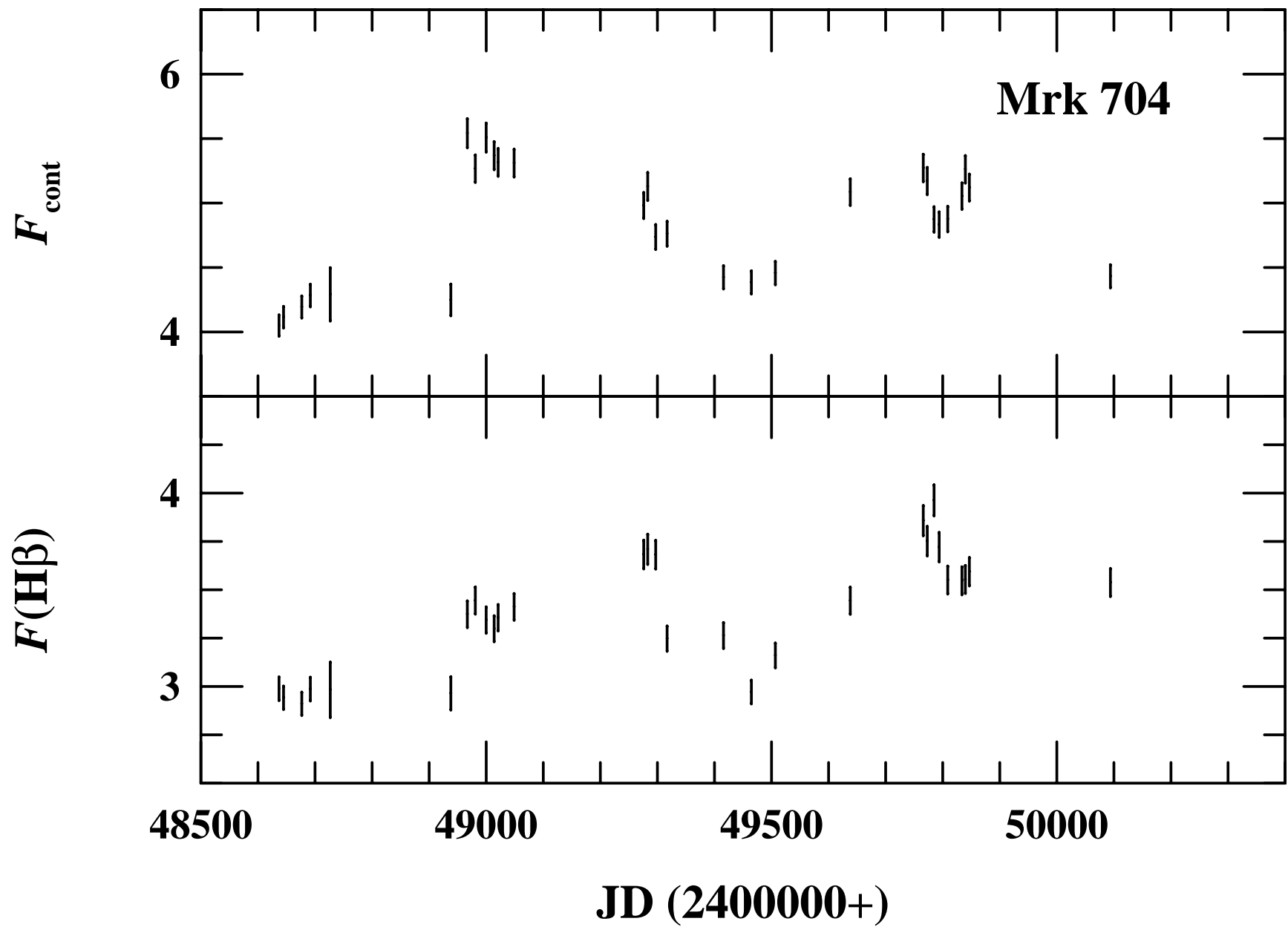
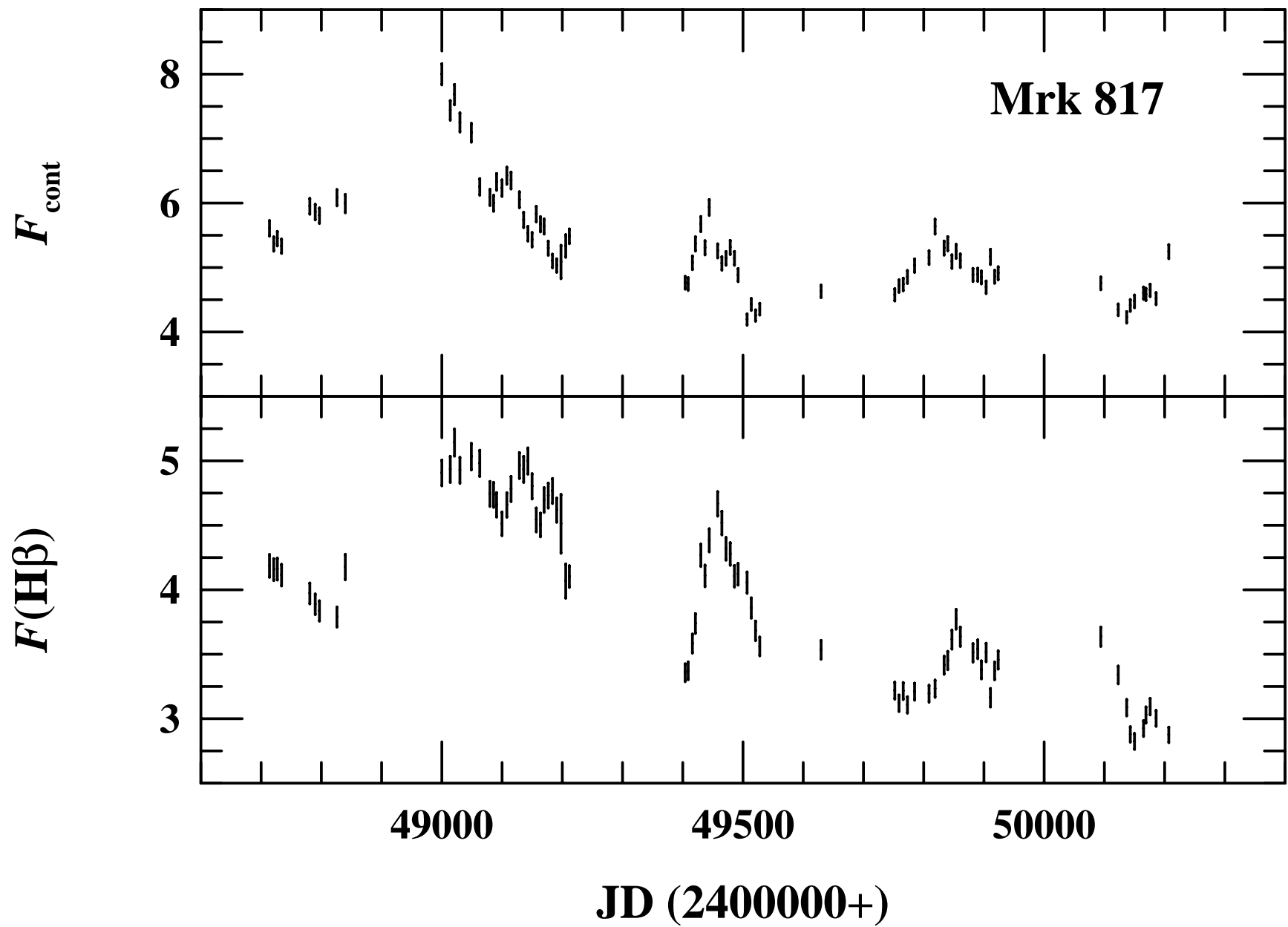
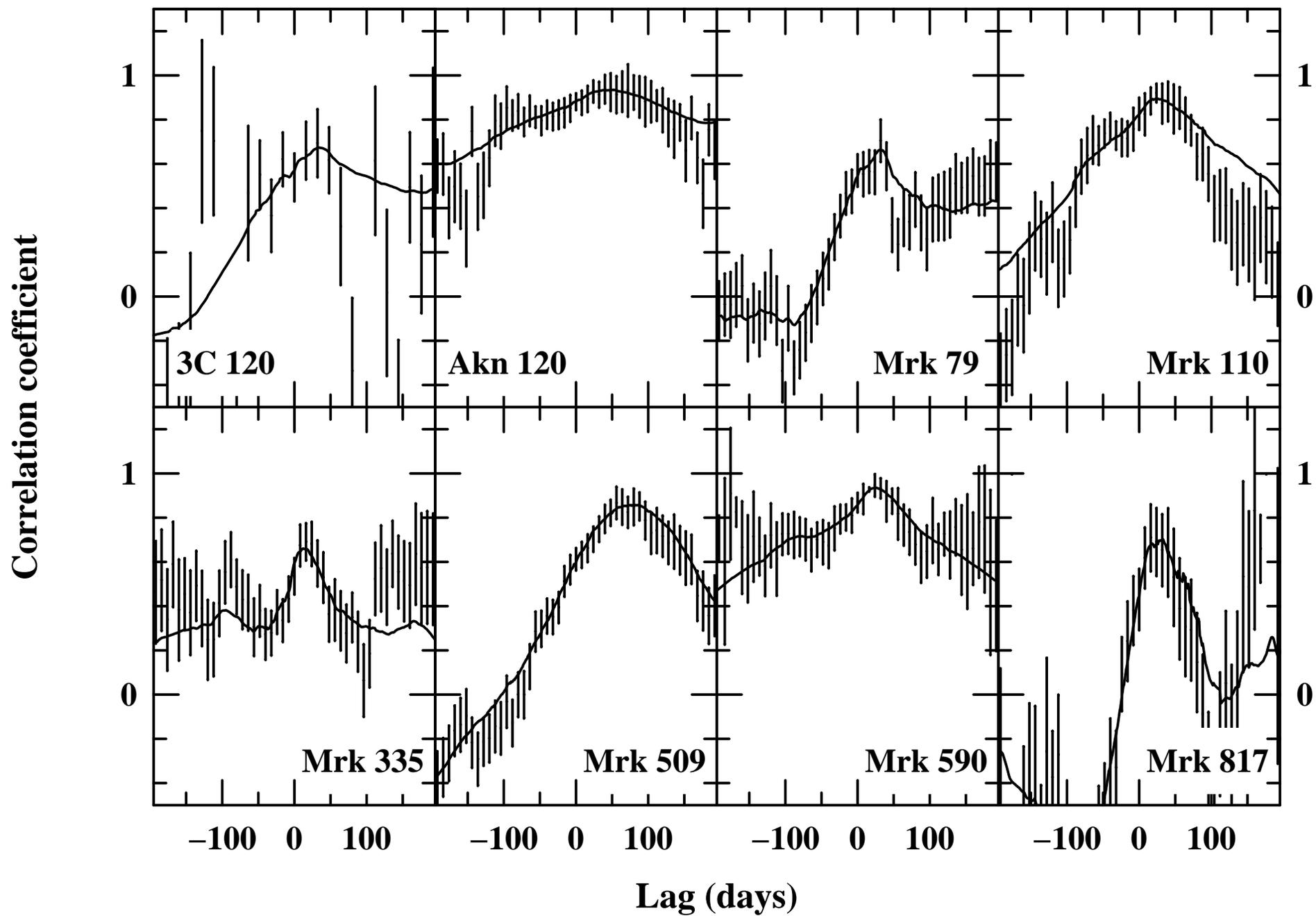
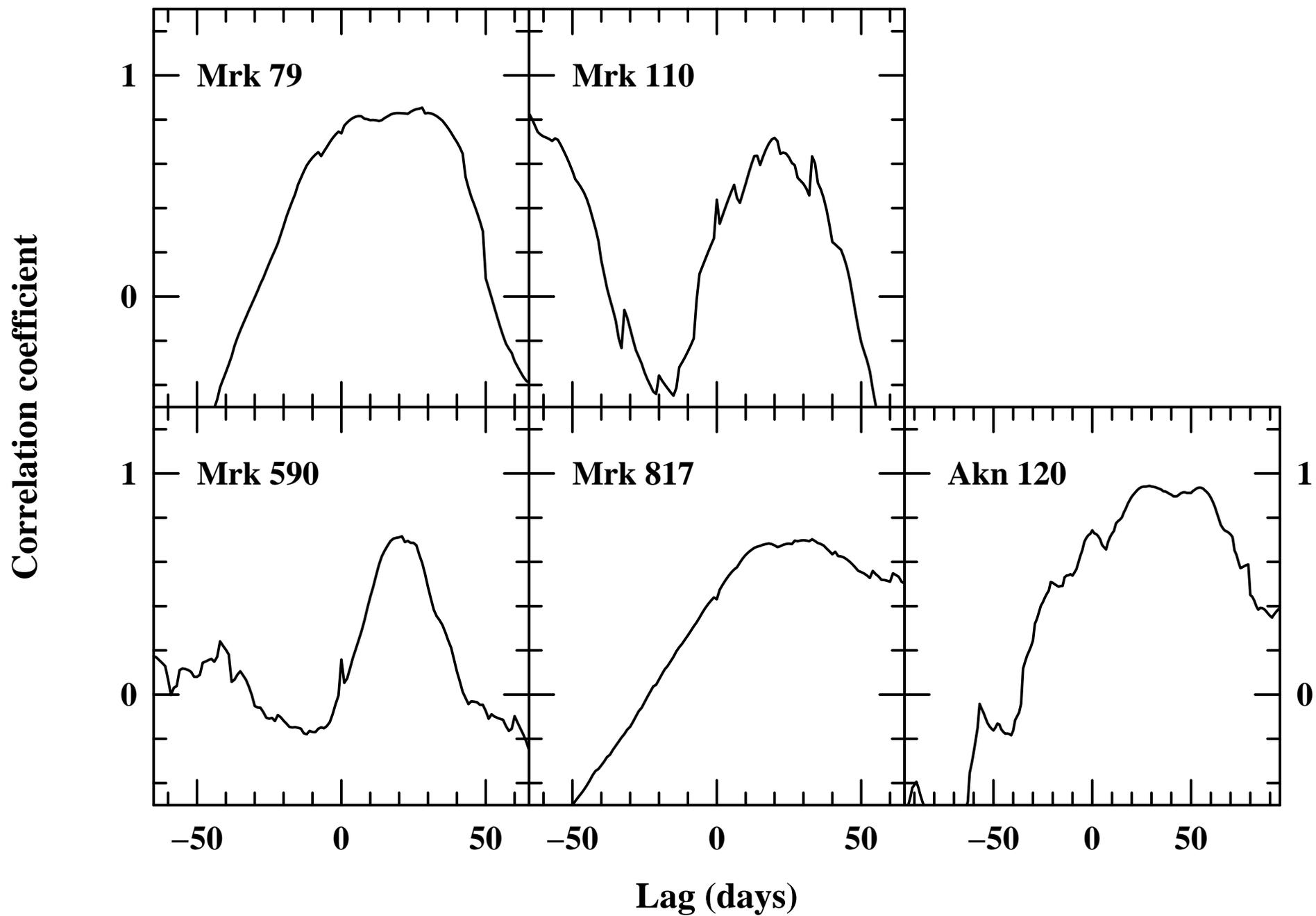
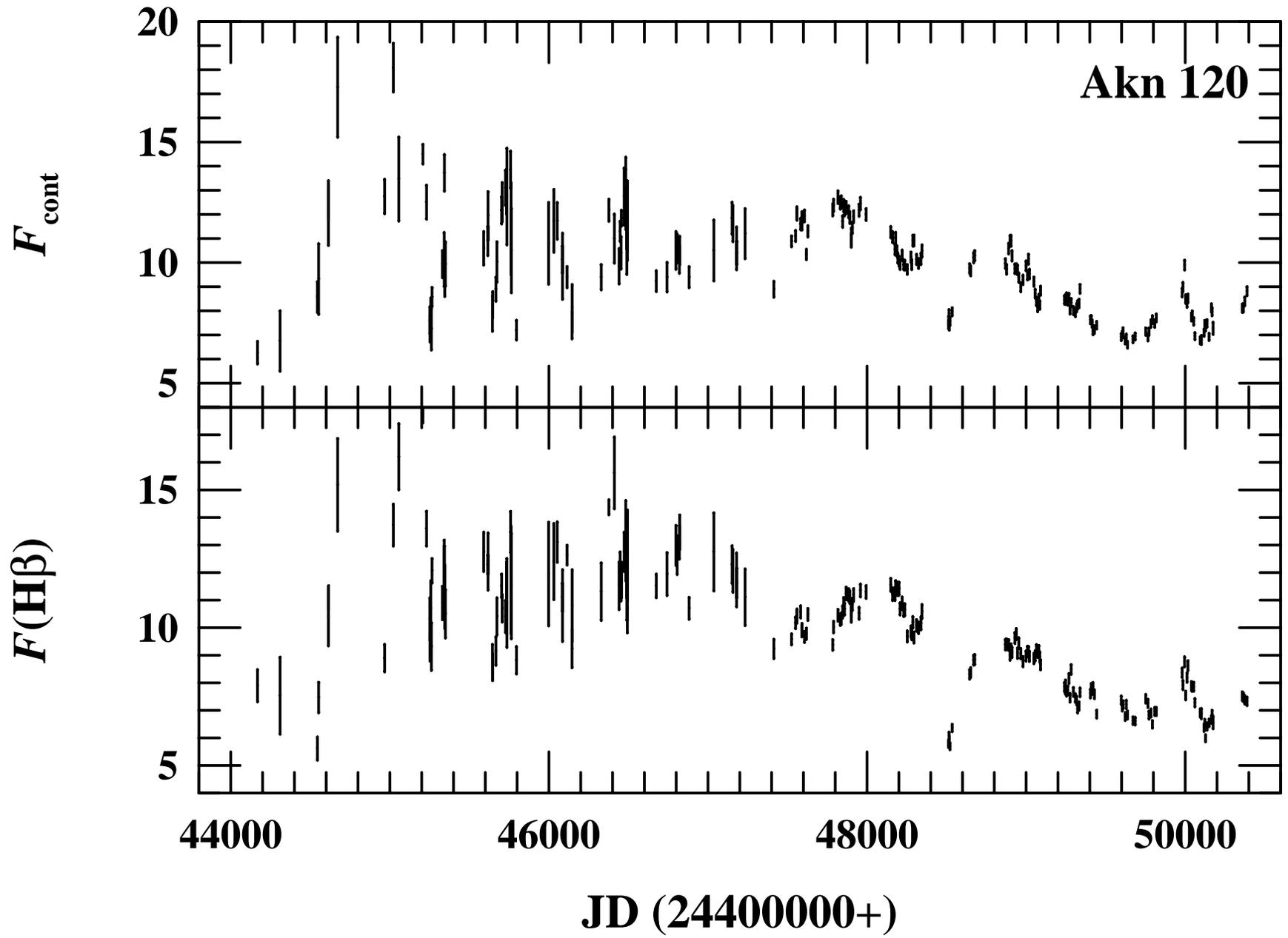
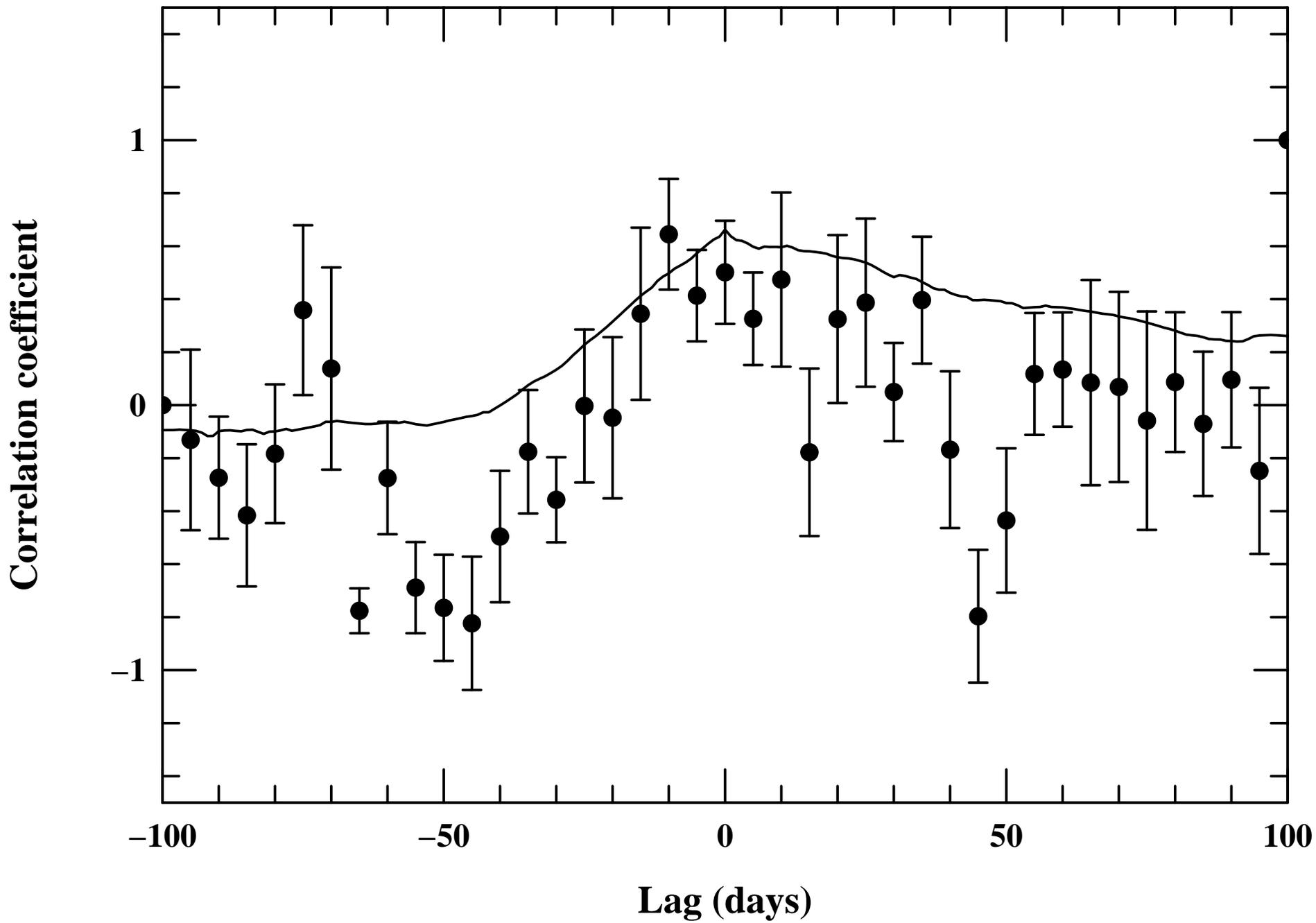
\end{document}